\documentclass[aps,prd,preprintnumbers, superscriptaddress,letterpaper,nofootinbib,twocolumn, longbibliography]{revtex4-2}
\usepackage[colorlinks=true,linkcolor=blue,citecolor=blue,urlcolor=blue]{hyperref}
\usepackage{lineno}
\usepackage{amssymb}
\usepackage{amsmath}
\usepackage{braket}
\usepackage{graphicx}
\usepackage{adjustbox}
\usepackage{array}
\usepackage{soul}

\begin{document}

\preprint{MIT-CTP/5791}

%\title{Signature of Axion-Like Particles in the Polarization of Gamma Ray from Pulsars}
\title{Searching for GeV Gamma-Ray Polarization and Axion-Like Particles with AMS-02}


\newcommand\AMSImit{       % Cambridge 
	Massachusetts Institute of Technology (MIT), Cambridge, Massachusetts 02139, USA}
	
\author{Xiuyuan Zhang}
\affiliation{\AMSImit}

\author{Yi Jia$^*$}
\affiliation{\AMSImit}

\author{Tracy R. Slatyer}
\affiliation{\AMSImit}
\affiliation{Department of Physics, Harvard University, Cambridge, Massachusetts 02138, USA}
\affiliation{Radcliffe Institute for Advanced Study at Harvard University, Cambridge,  Massachusetts 02138, USA}

%\thanks{}

\altaffiliation{Now at the Institute of High Energy Physics, Beijing.}


%\date{\today}

\begin{abstract}
We study the detectability of GeV-band gamma-ray polarization with the AMS-02 experiment and its proposed successor AMS-100, from Galactic and extragalactic sources. Characterizing gamma-ray polarization in this energy range could shed light on gamma-ray emission mechanisms in the sources; physics beyond the Standard Model, such as the presence of axion-like particles (ALPs), could also induce a distinctive energy-dependent polarization signal due to  propagation effects in magnetic fields. We present estimates for the minimum detectable polarization from bright sources and the forecast reach for axion-like particles (ALPs). We show that AMS-02 will have sensitivity to gamma-ray polarization only for the brightest steady-state Galactic sources, such as the Vela and Geminga pulsars; it is not expected to be capable of detecting polarization in Galactic or extragalactic sources that have been previously proposed as good targets for ALP searches with gamma-ray intensity measurements. However, AMS-100 observing the extragalactic source NGC1275 would be expected to probe new parameter space even for unfavorable B-field models, with prospects to measure the energy-dependence of such a signal.
%\hl{We show that for an example bright extragalactic source (NGC1275), AMS-02 will only have sensitivity to ALP-induced polarization from currently-open parameter space if the B-field configuration is unusually favorable to a signal. However, a successor experiment such as AMS-100 would be expected to probe new parameter space even for unfavorable B-field models, with prospects to measure the energy-dependence of such a signal.} 
For Galactic sources, polarization measurements could provide a unique test of scenarios where ALPs induce energy-dependent features in the photon intensity. However, in the absence of a bright transient source (such as a Galactic supernova), the parameter space that would be probed by this approach with ten years of AMS-100 data is already nominally excluded by other experiments, although this conflict may be avoided in specific ALP models.    
\end{abstract}

%\keywords{}

\maketitle


\section{Introduction}

The existence of dark matter has long been one of the greatest puzzles in our understanding of the Universe. 
 One of the leading candidates for dark matter particles is a hypothetical particle called the axion~\cite{Weinberg1978,Wilczek1978, Dine:1982ah, Abbott:1982af, Preskill:1982cy} introduced by Peccei and Quinn~\cite{Peccei1977,Peccei:1977ur} to resolve the puzzle of the lack of observed CP violation in quantum chromodynamics (QCD). More generally, axion-like particles (ALPs) are ultralight pseudoscalar bosons predicted in many theoretical frameworks beyond the Standard Model of particle physics, e.g.,~\cite{Svrcek:2006yi,Arvanitaki:2009fg,Cicoli:2012sz}. In contrast to QCD axions, the characteristic parameters of ALPs---the mass $m_{a}$ and the photon-axion coupling $g_{a\gamma\gamma}$---are not necessarily related to each other. In the presence of an external magnetic field, the photon-ALP mixing can lead to energy-dependent modifications to both the flux and the polarization of photons from astrophysical sources, over a wide range of frequencies. 

 The effects of ALPs on the polarization of light have been studied for various astrophysical sources (e.g., gamma-ray bursts~\cite{Bassan:2010ya},  magnetic white dwarfs~\cite{Gill:2011yp}, active galactic nuclei~\cite{Horns:2012pp,Day:2018ckv}, and pulsars~\cite{Liu:2019brz}). 
These studies are typically applied to photon energies below the MeV scale. 
Extending the studies of polarization to the GeV region requires the measurement of electron-positron pair production. 
To date, no data on polarization has been released for photons of energies above the 
pair creation threshold ($\sim$ 1 MeV), although the Fermi Large Area Telescope
(Fermi-LAT) has the potential to detect high degrees of linear polarization from some of the brightest gamma-ray sources~\cite{Giomi:2016brf}.
Challenges and future directions for gamma-ray polarimetry have been reviewed in Ref.~\cite{Ilie:2019yvs}.

In the gamma-ray band, ALP-induced modification of \emph{ intensity} spectra from Galactic pulsars~\cite{Majumdar:2018sbv} and supernova remnants~\cite{Xia:2018xbt} has already been studied in data from the Fermi-LAT gamma-ray telescope. Intriguingly, these studies have claimed a hint of a possible signal, with a favored mass of $\sim$ neV and coupling $g_{a\gamma\gamma} \sim 10^{-10}$ GeV$^{-1}$.  
 The tension of these results with the limits from the CAST helioscope~\cite{CAST:2017uph} might be reconciled by considering the different environmental conditions inside the Sun and the dilute interstellar medium where the photon-axion mixing occurs~\cite{Pallathadka:2020vwu}, or by non-standard ALP couplings \cite{Choi:2018mvk}. Additional experimental data such as polarization from these Galactic sources could provide valuable input in testing the ALP hypothesis and perhaps probing the properties of ALPs.

The Alpha Magnetic Spectrometer (AMS-02) has a high precision silicon tracker~\cite{Alpat:2010zz,Ambrosi:2017gez,Ambrosi:2015xyt} with which to measure the trajectories of electrons and positrons; it provides the opportunity for such a GeV-scale gamma-ray polarization measurement.
AMS-02 is a multi-purpose high energy particle experiment that has been installed in the International Space Station since 2011. Besides precision results on charged cosmic rays (see a review of AMS-02 results in Ref.~\cite{AMS:2021nhj}), AMS-02 has the capability to measure gamma rays in the MeV-GeV region for a wide range of astrophysical sources~\cite{Beischer:2020rts}, and its advantages for measuring gamma-ray polarization have been briefly discussed in Ref.~\cite{Bernard:2022jrj}.  

In this paper, we estimate the possible imprint of ALP dark matter in energy-dependent gamma-ray polarization, and discuss the implications for the sensitivity of AMS-02 and proposed next generation spectrometer experiments in space. Proposals for such experiments include ALADInO and AMS-100 ~\cite{Schael:2019lvx,Battiston2021}; we will focus on AMS-100, which aims to have near-$4\pi$ sr sky coverage, and geometric acceptance exceeding that of AMS-02 by roughly a factor of 1000. We consider the extragalactic source NGC1275 at the center of the Perseus cluster, which has been studied in the context of ALP searches using X-ray polarimetry (e.g.~in Ref.~\cite{Day:2018ckv}) and gamma-ray intensity measurements (e.g.~\cite{Fermi-LAT:2016nkz, Pallathadka:2020vwu, Cheng:2020bhr}), and a range of bright Galactic pulsars and supernovae. 

This paper is structured as follows: 
in section \ref{sec:Pheno}, we summarize the formalism to model photon-ALP mixing in our galaxy,   discuss the Galactic and extragalactic sources we consider, and describe our modeling of the relevant magnetic fields.
In section \ref{AMS}, we describe how polarization is measured by the AMS-02 detector. In section \ref{sensitivity}, we describe our pipeline for computing the projected sensitivity for the current AMS-02 and future proposed experiments. We present our results and discuss implications for future measurements in section \ref{sec:results}. Finally,  in section \ref{results}, we summarize our findings and discuss the prospects for detection of (ALP-induced) polarization with AMS-02 and successors. In the appendices, we discuss the dependence of our results on the choice of energy binning, and the sensitivity of our Galactic results to the inclusion of individual sources.

\section{\label{sec:Pheno}Predicting the ALP-induced polarization signal}
\subsection{\label{mixing}Photon-ALP mixing}
Photon-ALP mixing in the presence of an external magnetic field is described by the following Lagrangian:
\begin{linenomath}
\begin{equation}
\mathcal{L}
=g_{a\gamma\gamma}a\bf E \cdot \bf B,
\end{equation}
\end{linenomath}
where $g_{a\gamma\gamma}$ is the photon-axion coupling constant, $a$ is the ALP field with mass $m_{a}$, and $\bf E$ and $\bf B$ are the electric and magnetic fields, respectively. Assuming the variation of the magnetic field in space occurs on much larger scales than the photon or axion wavelength, the equations of motion for a photon or ALP of energy $E$ travelling along the $z$-direction can be reduced to a linearized form~\cite{Raffelt:1987im}:
\begin{linenomath}
\begin{equation}\label{eq2}
    (E + \mathcal{M} - i\partial_{z})
    \begin{pmatrix}
        \Ket{\gamma_x}\\
        \Ket{\gamma_y}\\
        \Ket{a}
    \end{pmatrix} = 0,
\end{equation}
\end{linenomath}
where $\Ket{\gamma_x}$ and $\Ket{\gamma_y}$ correspond to the two linear polarization states of the photon field. We can always choose a coordinate system such that the external magnetic field  transverse to the propagation direction, $B_{\perp}$,  aligns with the $y$-direction. The mixing matrix is then~\cite{Horns:2012kw,Meyer:2014epa} 
\begin{linenomath}
\begin{equation}\label{eq3}
\mathcal{M} = 
       \begin{pmatrix}
        \Delta_{\perp}    & 0                  & 0\\
        0                 & \Delta_{\parallel} & \Delta_{\gamma a}\\
        0                 & \Delta_{\gamma a} & \Delta_{a} 
        \end{pmatrix},
\end{equation}
\end{linenomath}
 where $\Delta_{\perp}= \Delta_{\text{pl}}+2\Delta_{\text{QED}}$, $\Delta_{\parallel}=\Delta_{\text{pl}}+7/2\Delta_{\text{QED}}$, the ALP mass term is  $\Delta_{a}=-m_{a}^2/(2E)$, and the ALP-photon mixing term is $\Delta_{\gamma a}=g_{a\gamma \gamma} B_{\perp}/2$.  The photon mass term $\Delta_{\text{pl}}=-\omega_{\text{pl}}^2/(2E)$ accounts for plasma effects during photon propagation, where   $\omega_{\text{pl}}=\sqrt{(4\pi\alpha N_e)/m_e}$ is the plasma frequency calculated from the fine structure constant $\alpha$, the electron number density $N_e$, and the electron mass $m_e$.  $\Delta_{\text{QED}}$ is an one-loop QED correction to the photon polarization in the presence of an external magnetic field. Note that off-diagonal terms that induce Faraday rotation are negligible \cite{Horns:2012kw}, and therefore are dropped within our treatment.  
 The numerical values for the parameters can be calculated (e.g., in Refs.~\cite{Horns:2012kw, Bassan:2010ya}) to be
\begin{linenomath}
\begin{equation}
    \begin{split}
        \Delta_{\text{pl}}&=-1.1\times 10^{-7}\left(\frac{N_e}{10^{-3}\text{cm}^{-3}}\right)\left(\frac{E}{\text{GeV}}\right)^{-1}\text{kpc}^{-1}, \\
        \Delta_{\text{QED}}&=4.1\times 10^{-9}\left(\frac{E}{\text{GeV}}\right)\left(\frac{B_{\perp}}{\mu \text{G}}\right)^2 \text{kpc}^{-1}, \\
        \Delta_{a}&=-7.8 \times 10^{-2} \left(\frac{m_a}{\text{neV}}\right)^2\left(\frac{E}{\text{GeV}}\right)^{-1}\text{kpc}^{-1}, \\
        \Delta_{\gamma a}&=1.52 \times 10^{-2} \left(\frac{g_{a \gamma \gamma}}{10^{-11}\text{GeV}^{-1}}\right)\left(\frac{B_{\perp}}{\mu \text{G}}\right)\text{kpc}^{-1}.\\
    \end{split}
\end{equation}
\end{linenomath}

The eigenvalues of the mixing matrix can be easily found to be $\Delta_0=\Delta_{\perp}$ and $\Delta_{\pm}=\left(\Delta_{\parallel}+\Delta_a \pm \sqrt{(\Delta_{\parallel}-\Delta_a)^2+4\Delta_{\gamma a}^2}\right)/2$. The solution is proportional to $\exp\left[-i(E+\Delta_{\pm})z\right]$. The oscillation wave number is thus $\Delta_\text{osc}=\sqrt{(\Delta_{\parallel}-\Delta_a)^2+4\Delta_{\gamma a}^2}$~\cite{Meyer:2014epa}. Since the first term inside the square root is inversely proportional to energy and the second term is independent of energy, the oscillation pattern will change at a critical energy $E_c$ when the two terms are around the same order $(\Delta_{\parallel}-\Delta_a)^2 \sim 4\Delta_{\gamma a}^2$. We can solve for the critical energy as:
\begin{linenomath}
\begin{equation}
    \begin{split}
        E_c &\simeq \frac{1}{2} |m_a^2-\omega_\text{pl}^2|B_{\perp}^{-1}g_{a\gamma \gamma}^{-1}\\
        &=2.5\text{GeV} \frac{|m_a^2-\omega_\text{pl}^2|}{(1\text{neV})^2}\left(\frac{B_{\perp}}{\mu \text{G}}\right)^{-1}\left(\frac{g_{a\gamma \gamma}}{10^{-11}\text{GeV}^{-1}}\right)^{-1}. 
    \end{split}
\end{equation}
\end{linenomath}

At energies above $E_{c}$, the oscillation wave number $\Delta_\text{osc}$ is dominated by the energy independent part $\Delta_{\gamma a}$. The conversion probability for photon mixing into ALP after a distance $d$ can be found to be~\cite{Horns:2012kw}
\begin{linenomath}
\begin{equation}
        P_{\gamma a}^{(0)}=(\Delta_{\gamma a} d)^2\frac{\sin^2(\Delta_\text{osc} d/2)}{(\Delta_\text{osc}d/2)^2},
\end{equation}
\end{linenomath}
and so this probability becomes maximal and energy-independent at energies above $E_{c}$. At lower energies, the oscillation probability is suppressed by $(\Delta_{\gamma a}/\Delta_{\text{osc}})^2$.

The oscillation length~\cite{Majumdar:2018sbv} scales as the inverse of the oscillation wave number. Thus, if we can treat the magnetic field $B_\perp$ as being roughly constant, the oscillation length can be written as
\begin{linenomath}
\begin{equation}
        l_{\text{osc}}=32\text{kpc} \sqrt{1+(E_c/E)^2}\left(\frac{B_\perp}{\mu \text{G}}\right)^{-1}\left(\frac{g_{a \gamma \gamma}}{10^{-11}\text{GeV}^{-1}}\right)^{-1}. 
\end{equation}
\end{linenomath}
We observe that, for typical values of the magnetic field in our Galaxy ($\mathcal{O}(\mu G)$, and for $E$ comparable to $E_c$, oscillations are relevant for sources at Galactic-scale distances for $g_{a\gamma \gamma} \sim 10^{-11}$ GeV$^{-1}$. 

All else being equal, a larger source distance $d$ allows for more oscillation and thus is favorable for probing larger $l_\text{osc}$, corresponding to smaller couplings $g_{a\gamma \gamma}$. 
Thus extragalactic sources such as galaxy clusters can allow access to smaller couplings, as can strong fields in the neighborhood of the sources themselves. However, our knowledge of the magnetic field along the line of sight for extragalactic sources is generally rather limited compared to our knowledge of the magnetic field of our Galaxy. Thus for each extragalactic source, the magnetic field needs to be modeled independently (as done in e.g.~\cite{Day:2018ckv}). We will work out one example for an extragalactic source (NGC1275) that would allow sensitivity to smaller $g_{a\gamma\gamma}$ for plausible magnetic field models. 

In order to study the mixing of photons from an unpolarized source, we need to work with the density matrix instead~\cite{Horns:2012kw}:
\begin{linenomath}
\begin{equation}
    \rho=\begin{pmatrix}
        \Ket{\gamma_x}\\
        \Ket{\gamma_y}\\
        \Ket{a}
    \end{pmatrix}\otimes (\bra{\gamma_x}, \bra{\gamma_y}, \bra{a}),
\end{equation}
\end{linenomath}
that obeys the von-Neumann-like equation
\begin{linenomath}
\begin{equation}
        i\frac{d\rho}{dz}=[\rho, \mathcal{M}_0]. 
\end{equation}
\end{linenomath}
This equation can be solved using the transfer matrix $\mathcal{T}$, defined as the solution to Eq.~\ref{eq2} with the initial condition $\mathcal{T}(0, 0;E)=1$. We can write $\rho(z)=\mathcal{T}(z, 0; E)\rho(0)\mathcal{T}^{\dagger}(z, 0; E)$. In general, $B_{\perp}$ will not remain in the same direction along the line of sight, and might form an angle $\psi$ with the $y$-direction that changes along the $z$-direction. If we divide the distance from the source to the Earth into $N$ consecutive steps and assume the angle $\psi$ does not change within each step, then the propagation of the whole beam can be described by the transfer matrix 
\begin{linenomath}
\begin{equation}
        \mathcal{T}(z_N, z_1; \psi_N, ..., \psi_1;E)=\prod_i^N\mathcal{T}(z_{i+1}, z_i;\psi_i;E).
\end{equation}
\end{linenomath}

Finally, let us estimate how our signal sensitivity will depend on the axion parameters $m_a$ and $g_{a\gamma \gamma}$. We will be interested both in searching for the presence of an induced polarization signal, and (if such a signal is detected) measuring its energy dependence to constrain its origin. For the second type of search, we require the polarization to vary with energy in the GeV band. For the first type of search,  $E \gg E_c$ is favorable as this maximizes the conversion probability. However, the energy-dependence of the conversion probability is most pronounced when  $E_c$ is greater than the energies of interest (i.e.~the AMS-02 energy band around 0.1-10 GeV). 

Note that we can rewrite the conversion probability in terms of the critical energy $E_c$ and oscillation length $l_{\text{osc}}$:
\begin{linenomath}
\begin{equation}
        P_{\gamma a}^{(0)}=\frac{\sin^2(2\pi d/l_{\text{osc}})}{((E_c/E)^2+1)}.
        \label{eq:convprob}
\end{equation}
\end{linenomath}

Thus we see that when $E \gg E_c$, there is only one relevant scale $l_{\text{osc}}$. The oscillation signal will then generally be non-negligible once the distance to the source $d$ becomes comparable to $l_{\text{osc}}$. Furthermore, when $E \gg E_c$, 
$l_{\text{osc}} \propto 1/g_{a\gamma \gamma}$. Thus we expect to lose sensitivity to the oscillation signal at a fixed value of $g_{a\gamma \gamma}$ for each source, corresponding to $l_{\text{osc}}$ becoming  large compared to $d$ and thus ensuring a small conversion probability even for $E\gg E_c$, if the approximation $E \gg E_c$ is good in the region that is marginally constrained. 

This argument holds for all axion masses, but for $m_a \gg \omega_\text{pl}$, $E_c$ increases proportionally to $m_a^2/g_{a\gamma \gamma}$. Thus a line with $g_{a\gamma \gamma} \propto m_a^2$ (starting at a critical mass determined by the plasma frequency) will keep $E_c$ fixed, and a line with this slope will form the boundary between the regions with $E \gg E_c$ (higher $g_{a\gamma \gamma}$) and $E \ll E_c$ (lower $g_{a\gamma \gamma}$). If we can estimate the polarization signal region as requiring $E \gtrsim E_c$ and $l_\text{osc} \gtrsim d$, then this will define a minimum testable $g_{a\gamma \gamma}$ which is mass-independent at low $m_a$ (fixed by $l_\text{osc} \lesssim d$) and then grows proportionally to $m_a^2$ at higher $m_a$ (where extrapolating the low-$m_a$ line would lead to $E \lesssim E_c$).

Now the energy variation in the polarization is suppressed in the region with $E \gg E_c$, but if $E \ll E_c$ leads to an undetectably small polarization signal overall, we might expect that $E_c$ must lie within some specific range of energies (comparable to those we observe) in order for the energy variation to be measurable. In this case, the sensitivity region will be bounded below in $g_{a\gamma \gamma}$ by the requirement that $l_\text{osc} \lesssim d$, and to the sides by $g_{a\gamma \gamma} \propto m_a^2$ lines of fixed $E_c$. The sensitivity region could potentially be improved in this case by combining gamma-ray polarization measurements with those from lower-energy instruments, but we will not pursue this idea further in this work.

We can refine this estimate slightly by noting that if $l_\text{osc} \ll d$, then small energy-dependent changes in $l_\text{osc}$ can affect the phase at the $\mathcal{O}(1)$ level in Eq.~\ref{eq:convprob}. For $E \gg E_c$, we can approximate the change in $l_\text{osc}$ over an energy range $\Delta E$ as:
\begin{align} \Delta l_\text{osc} \approx - \frac{E_c^2}{E^3} l_\text{osc} \Delta E \end{align}
Then $\Delta(d/l_\text{osc}) \approx (d/l_\text{osc}) (1 + \frac{E_c^2}{E^2})$ for $\Delta E \sim E$, and so the phase difference scales as $E_c^2/l_\text{osc} \propto g_{a\gamma \gamma} m_a^4/g_{a\gamma\gamma}^2 = m_a^4/g_{a\gamma \gamma}$. Thus we expect to lose sensitivity to the energy-variation signal at low ALP masses, as previously, but with a boundary curve of slope $g_{a\gamma \gamma} \propto m_a^4$ rather than $m_a^2$. We will see that these estimates are sufficient to understand the shape of our sensitivity forecasts.

These arguments also clarify how measurements of gamma-ray polarization could improve on previous studies in the X-ray band; detection of polarization is easiest for $E \gtrsim E_c$, with detection of energy dependence requiring $E\sim E_c$, so measuring polarization at higher $E$ gives access to parameter space that predicts higher values of $E_c$. Since where the plasma mass can be neglected, $E_c\propto m_a^2/g_{a\gamma \gamma}$, this implies we will be able to test higher $m_a$ for a given value of $g_{a\gamma \gamma}$; the high-$m_a$ cutoff associated with the critical energy will move to the right, proportionally to $\sqrt{E}$. Thus we expect that moving from keV-scale X-ray polarization to GeV-scale gamma-ray polarization could buy roughly three orders of magnitude of reach in $m_a$. Thus for example Ref.~\cite{Day:2018ckv} forecasts that X-ray polarization studies of the bright extragalactic source NGC1275 will have sensitivity down to $g_{a\gamma\gamma}$ slightly below $10^{-12} \text{GeV}^{-1}$, up to masses around $m_a\sim 10^{-12}$ eV (where there is a sharp cutoff in sensitivity); we may hope for a comparable sensitivity up to $m_a\sim 10^{-9}$ eV, although the sensitivity to $g_{a\gamma \gamma}$ will depend on our ability to measure small polarization fractions and hence on the photon statistics.

\subsection{\label{GMF}Modeling the Galactic magnetic field}

In this work, we model the Galactic magnetic field (GMF) as described in Ref.~\cite{Jansson2012}, with an updated set of parameters from Ref.~\cite{PlanckGMF2016}. For Galactic sources, we assume the axion-photon conversion takes place during propagation through the GMF (as opposed to conversion in the source), so our GMF model is relevant to the signal from all Galactic sources. This GMF model contains four different components: a disk component, a toroidal component, a out-of-plane component and a striated random field. The striated random field is aligned with the regular field everywhere but with its direction randomized at small scale; its strength can be modeled as the strength of the regular field with an extra $\mathcal{O}(1)$ multiplicative factor. So the combined field will only differ from the regular field by a similar multiplicative factor. 
 Therefore we only consider the disk, out-of-plane, and toroidal components; the disk and out-of-plane components generally dominate along the lines of sight to the sources we consider. The disk component lies strictly on the Galactic plane  and contains eight spiral arms with different strengths that decrease exponentially as they rotate away from the Galactic Center.\footnote{We note that, in the original paper~\cite{Jansson2012}, the stated expression for the dividing line between the spiral arms $r=r_{-x}\exp\left[\phi \tan(90^{\circ}-i)\right]$ should be $r=r_{-x}\exp\left[(\phi-\pi )\tan(i)\right]$ as noted in Ref.~\cite{2019ApJ...877...76K}.} The disk component slowly transits into the toroidal component as it deviates away from the Galactic plane. The out-of-plane component has a radial $B$-field contribution and a contribution perpendicular to the Galactic plane.

 \subsection{\label{sources}Sources of gamma rays}

In this work, we will consider both polarized and unpolarized sources of GeV-scale gamma rays. Polarization has been measured for a number of Galactic and extragalactic source classes across a broad range of frequencies, but measurements in the GeV band are challenging both because of a lack of instruments designed for polarimetry, and because of the relatively small photon counts from typical sources in this energy range. A review of results in the hard X-ray and soft gamma-ray band, and future prospects in gamma-rays, can be found in Ref.~\cite{Ilie:2019yvs}; the expected GeV-band sensitivity to polarization for the Fermi-LAT, which uses a similar detection technique to AMS-02 in this regard, has been estimated in \cite{Giomi:2016brf}.

Pulsars frequently have observed emission extending into the GeV band, which can be attributed to synchrotron radiation, curvature radiation, and/or inverse Compton scattering (e.g. \cite{Harding:2017ypk} and references therein). All of these mechanisms are capable of generating linearly polarized gamma-ray emission, but the degree of polarization and its energy dependence can potentially distinguish between them. Ref.~\cite{Harding:2017ypk} models the emission from the outer magnetosphere of rotation-powered pulsars, assuming gamma-ray emission comes from synchrotron and/or curvature radiation from accelerated primary electrons. That work predicts phase-averaged polarization degrees up to 40-60\% if curvature radiation is the dominant gamma-ray emission mechanism, or 10-20\% if synchrotron radiation is responsible, at GeV-band energies and above. One particularly bright pulsar and standard calibration source is the Crab pulsar/nebula. Multiple instruments have sought to measure polarized soft gamma-ray emission from the Crab in the keV-MeV band, finding central values for the linear polarization fraction ranging from around 15\% to nearly 50\% (e.g. \cite{Li:2022qei} and references therein).

It therefore seems plausible that pulsars may emit GeV-band gamma rays with a wide range of polarization fractions, so we test the effect of assuming either unpolarized or linearly polarized gamma-ray emission from pulsars at the source. The detection of polarized emission from pulsars would not (in itself) be a reliable indicator of the presence of ALPs, but an upper limit on polarization across multiple sources could still be used to set a bound on ALP-photon oscillation, and the energy dependence of the polarization could (in principle) be used to separate ALP-sourced polarization changes from an astrophysical background. The detection of polarized GeV-band gamma rays from Galactic pulsars would be an exciting discovery in itself, with implications for pulsar emission modeling, independent of any implications for ALP models.

Gamma-ray emission from supernova remnants may be hadronic or leptonic in origin. The hadronic gamma-ray emission from collisions of accelerated protons producing neutral pions (which then decay to gamma rays) is expected to be unpolarized, but inverse Compton signals may have a non-zero linear polarization fraction. In the X-ray band, polarized emission attributed to synchrotron has been detected from supernova remnants (e.g. \cite{Zhou:2023hwe}, who find a polarization degree around 20\%), and for soft gamma rays, Compton scattering has been discussed as a source of polarization \cite{Churazov:2018gxp}. Given the relatively low observed polarized fractions in the X-ray band and the likelihood that $\pi^0$ emission contributes in the GeV band (e.g.~\cite{2011ApJ...742L..30G}), we will assume that GeV-band emission from supernova remnants is predominantly unpolarized in this work (this is consistent with \cite{Giomi:2016brf} which considered only pulsar and blazar sources), although we would caution  against interpreting any detection of a modest polarization signal as requiring ALPs or other new physics.

Extragalactic gamma-ray sources may also have substantial polarization; e.g. Ref.~\cite{Zhang:2013bna} found that blazar polarization could reach maximum values as high as 70\% in the case of hadronic emission and 40\% for leptonic emission (although these are upper limits and the true values could be much lower). Again the physical mechanism generating the polarization in these cases is synchrotron radiation, including synchrotron from protons in the hadronic case. Ref.~\cite{Day:2018ckv} cites estimated linear polarization fractions of $0-5\%$ for NGC1275 in the X-ray band, and treats the linear polarization fraction as independent of energy.

Finally, gamma-ray emission could in principle also be circularly polarized at the source, but we expect that a large circularly polarized fraction would require some new mechanism for emission or scattering of the gamma rays (e.g. \cite{Boehm:2017nrl,Boehm:2019yit}). We will show some results for a circularly polarized source for completeness; note however that AMS-02's method for measuring gamma-ray polarization is only sensitive to the linear polarization fraction.

\subsection{\label{sources-gal}Selecting and modeling Galactic sources}

In this work, we will assume supernova remnants are unpolarized sources, and consider pulsars both for the polarized and unpolarized cases. We employ the same sample of six pulsars used in Ref.~\cite{Majumdar:2018sbv}. Note that in order to obtain a large oscillation signal, it is favorable to have a large $B_\perp$ component and a long path length. We observe that the radial component of the total $B$-field is relatively small compared with the azimuthal and perpendicular components. Accordingly, we prefer sources at low Galactic latitudes where the emitted photons can pass through the spiral arms, and sources that have relatively large and well-determined distances.
The sample we use corresponds to six bright pulsars selected according to these criteria. We also selected four of the five brightest supernova remnants from the Fermi-LAT fourth source catalog (4FGL)~\cite{Fermi-LAT:2019yla} using the same criteria; we did not include the W28 supernova remnant, which is the fourth brightest overall, due to a relatively small transverse magnetic field along the line of sight to this source. The distances to these sources are taken from Refs.~\cite{Xia:2018xbt, Acero_2016}. In addition to these sources, which are chosen to be favorable for observations of ALP-photon oscillation, we also consider two sources chosen purely for their brightness: the Vela and Geminga pulsars. These pulsars are very nearby and so are unlikely to be good candidates for observing ALP-induced polarization, but their high photon fluxes make them good candidates for observations of intrinsic (astrophysical) polarization.\footnote{The uncertainties on the distances to IC443 and W44 are not clearly defined in the original references (although they are certainly non-negligible). The distance to W51C is taken to be 5.5kpc, following Ref.~\cite{Xia:2018xbt}, which is also consistent with the measured value of $5.41^{+0.31}_{-0.28}$kpc given in Ref.~\cite{Sato_2010}. }

\begin{figure}[htbp]
\begin{center}
  \includegraphics[width=1.0\linewidth]{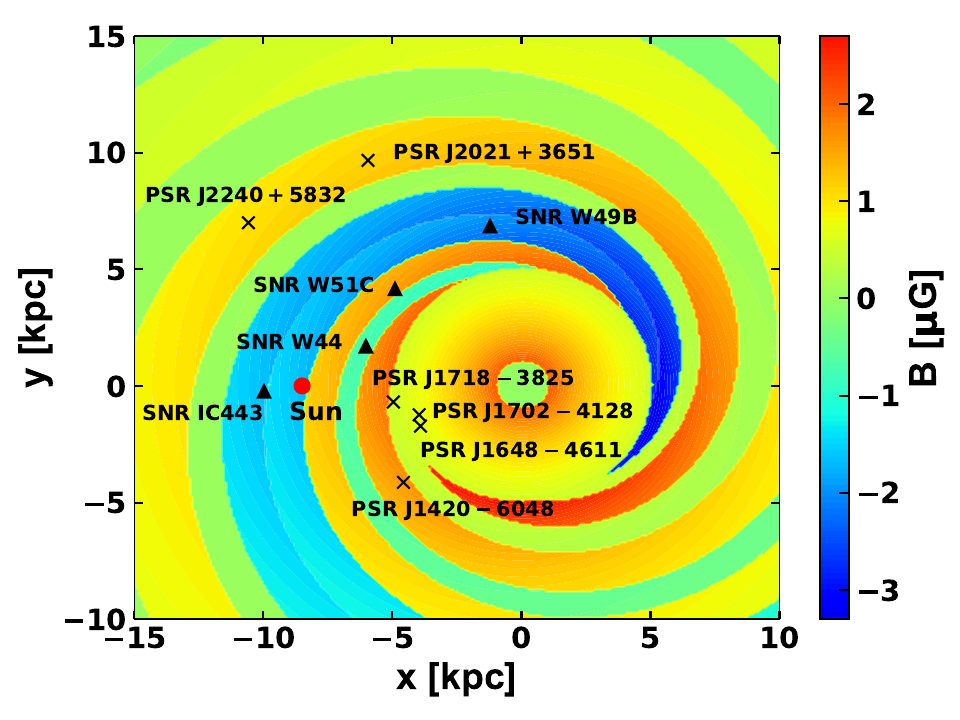}
 \caption{The strength of the Galactic magnetic field, based on the model in Ref.~\cite{Jansson2012}, as a function of position in the Galactic plane. The selected pulsars and supernova remnants are indicated as crosses and triangles, respectively. The position of the Sun is marked as a red point. Geminga and Vela are not indicated because of their proximity to the Sun (which also makes them poor targets for an ALP search).}
  \label{fig1}
 \end{center}
 \end{figure}

The positions of these sources are listed in Table~\ref{table1} and are shown, together with the magnetic field strength in the Galactic plane, in Fig.~\ref{fig1} (positions for Geminga and Vela are not shown because they essentially coincide with the Sun's location on the scale of this plot). The exposure for AMS-02~\cite{Beischer:2020rts} can be reduced in regions such as the North Pole and the South Atlantic Anomaly due to a decrease in data acquisition efficiency. The sources considered in this paper have been verified to be located outside these regions of low exposure.

  \begin{table}[htbp]
  \caption{The list of selected Galactic sources, with their Galactic longitude ($l$), latitude ($b$), and heliocentric distance ($d$) with corresponding errors. The first six entries correspond to pulsars and the others to supernovae (see the text for references).}
  \begin{tabular}{|m{0.2\linewidth}|m{0.2\linewidth}|m{0.2\linewidth}|m{0.2\linewidth}|}
  \hline
   \centering Source name & \centering $l$[$^{\circ}$] & \centering $b$[$^{\circ}$] & \centering $d$ [kpc] \tabularnewline
  \hline
  \centering J1420-6048 & \centering 313.54 & \centering 0.23 & \centering 5.7$\pm$ 0.9 \tabularnewline
   \centering J1648-4611 & \centering 339.44 & \centering -0.79 & \centering 4.9 $\pm$ 0.7 \tabularnewline
     \centering J1702-4128 & \centering 344.74 & \centering 0.12 &  \centering 4.7 $\pm$ 0.6  \tabularnewline
     \centering J1718-3825 & \centering 348.95 & \centering -0.43 & \centering 3.6 $\pm$ 0.4 \tabularnewline
  \centering J2021+3651 & \centering 75.22 & \centering 0.11 & \centering $10^{+2}_{-4}$ \tabularnewline   
   \centering J2240+5832 & \centering 106.57 & \centering -0.11 & \centering 7.3 $\pm$ 0.7 \tabularnewline
   \centering IC443 & \centering 189.065 & \centering  3.235 & \centering 1.5 \tabularnewline   
    \centering W44 & \centering 34.560 & \centering  -0.497 & \centering 3 \tabularnewline   
    \centering W51C & \centering 49.131 & \centering  -0.467 & \centering 5.5 \tabularnewline     
    \centering W49B & \centering 43.2515 & \centering  -0.1761 & \centering $10\pm 2$ \tabularnewline  
    \centering Geminga & \centering 195.13 & \centering  4.27 & \centering 0.25 \tabularnewline 
    \centering Vela & \centering 263.55 & \centering  -2.79 & \centering 0.29 \tabularnewline  
     \hline    
  \end{tabular}
  \label{table1}
  \end{table}

The pulsars are fairly young and rotation powered. Their emission spectra are modeled as a power law with an exponential cutoff as in Ref.~\cite{Majumdar:2018sbv}:
\begin{linenomath}
\begin{equation}\label{eq7}
  \frac{dN}{dE}=N_0\left(\frac{E}{E_0}\right)^{-\Gamma}\exp\left[-\frac{E}{E_\text{cut}}\right].
\end{equation}
\end{linenomath}

The parameters include the normalization factor $N_0$ at scale energy $E_0$, photon index $\Gamma$, and cutoff energy $E_\text{cut}$. Values for these parameters (taken from Ref.~\cite{Majumdar:2018sbv}) are presented in Table~\ref{table2}. 

  \begin{table}[htbp]
  \caption{Best fit spectral parameters for pulsars, with combined uncertainties (shown in parentheses), taken from Ref.~\cite{Majumdar:2018sbv}, including the normalization factor $N_0$ at scale energy $E_0$, photon index $\Gamma$, and cutoff energy $E_\text{cut}$.}
  \begin{tabular}{|m{0.2\linewidth}|m{0.3\linewidth}|m{0.125\linewidth}|m{0.15\linewidth}|m{0.125\linewidth}|}
  \hline
   \centering Pulsar name & \centering $N_0$\\\scalebox{0.8}{$ [10^{-9}\rm{MeV}^{-1}\rm{cm}^{-2}\rm{s}^{-1}]$} & \centering $E_0$\\\scalebox{0.8}{$[\rm{GeV}]$} & \centering $\Gamma$ & \centering $E_\text{cut}$\\\scalebox{0.8}{$[\rm{GeV}]$}\tabularnewline
  \hline
  \centering J1420-6048 & \centering 0.0014(2) & \centering 5.6 & \centering 1.79(4) & \centering 4.3(4) \tabularnewline
   \centering J1648-4611 & \centering 0.0022(1) & \centering 2.9 & \centering 0.98(3) & \centering 3.1(2)\tabularnewline
     \centering J1702-4128 & \centering 0.15(3) & \centering 0.1 &  \centering 0.8(1) & \centering 0.8(1) \tabularnewline
     \centering J1718-3825 & \centering 0.021(1) & \centering 1.2 & \centering 1.58(4) & \centering 2.2(2) \tabularnewline
  \centering J2021+3651 & \centering 0.15(1) & \centering 0.8 & \centering 1.59(3) & \centering 3.2(3) \tabularnewline   
   \centering J2240+5832 & \centering 0.0065(1) & \centering 1.2 & \centering 1.5(1) & \centering 1.6(4) \tabularnewline
     \hline    
  \end{tabular}
  \label{table2}
  \end{table}

  The emission spectra of the supernova remnants are modeled by a log-normal representation as in the Fourth Fermi-LAT catalog~\cite{Fermi-LAT:2019yla},

 \begin{linenomath}
\begin{equation}\label{eq13}
  \frac{dN}{dE}=K\left(\frac{E}{E_0}\right)^{-\alpha -\beta \log(E/E_0)}
\end{equation}
\end{linenomath} 
 with the best fit values of $K, E_0, \alpha, \beta$ given in Table~\ref{table3}. 

  \begin{table}[htbp]
  \caption{Best fit spectral parameters for supernova remnants and the extragalactic source NGC1275, from the Fourth Fermi-LAT catalog \cite{Fermi-LAT:2019yla}; see Eq.~\ref{eq13} for parameter definitions.}
  \begin{tabular}{|m{0.2\linewidth}|m{0.3\linewidth}|m{0.14\linewidth}|m{0.13\linewidth}|m{0.13\linewidth}|}
  \hline
   \centering SNR name & \centering $K$\\\scalebox{0.8}{$ [10^{-9}\rm{MeV}^{-1}\rm{cm}^{-2}\rm{s}^{-1}]$} & \centering $E_0$\\\scalebox{0.8}{$[\rm{GeV}]$} & \centering $\alpha$ & \centering $\beta$\tabularnewline
  \hline
  \centering IC443 & \centering 0.0025754 & \centering 4.55086 & \centering 2.2838 & \centering 0.1226 \tabularnewline
   \centering W44 & \centering 0.0080814 & \centering 2.79088 & \centering 2.5268 & \centering 0.2389\tabularnewline
     \centering W51C & \centering 0.0050819 & \centering 2.76802 &  \centering 2.2054 & \centering 0.1086 \tabularnewline
     \centering W49B & \centering 0.00077392 & \centering 4.55187 & \centering 2.2827 & \centering 0.1118\tabularnewline
     \centering NGC1275 & \centering 0.039039 & \centering 0.9749 & \centering 2.0594 & \centering 0.0719\tabularnewline
     \hline    
  \end{tabular}
  \label{table3}
  \end{table}

Unlike the other pulsars, Geminga and Vela have significantly curved spectra and are described by a sub-exponentially cutoff power law,
\begin{linenomath}
\begin{equation}\label{eq7}
  \frac{dN}{dE}=N_0\left(\frac{E}{E_0}\right)^{\gamma_0+d/b}\exp\left[\frac{d}{b^2}\left(1-\left(\frac{E}{E_\text{cut}}\right)^b\right)\right].
\end{equation}
\end{linenomath}

For Geminga, we have a prefactor $N_0=0.30696 \times 10^{-9}\rm{MeV}^{-1} \rm{cm}^{-2}\rm{s}^{-1}$, local spectral index $\gamma_0=-2.0288$, local curvature $d=0.71448$, index $b=0.6806$, and energy scale $E_0=1.70533$~GeV. For Vela, we have $N_0=0.41083 \times 10^{-9}\rm{MeV}^{-1} \rm{cm}^{-2}\rm{s}^{-1}$, local spectral index $\gamma_0=-2.2607$, local curvature $d=0.57942$, index $b=0.4922$, and energy scale $E_0=1.97637$~GeV. 
 \subsection{\label{Extragalactic}Selecting and modeling extragalactic sources}

As mentioned in section \ref{sec:Pheno}, distant sources should generically allow us to probe lower couplings, especially if we have reason to believe there is a substantial magnetic field along the line of sight.
Most of the prominent extragalactic gamma ray sources are active Galactic nuclei (AGNs), the brightest of which have comparable gamma ray flux to bright Galactic sources over the energy range of AMS-like detectors. Examples of such sources include NGC1275 and Markarian 421~\cite{Fermi-LAT:2019yla}. 
Furthermore, one of the most well studied AGNs, NGC1275, is at the center of the Perseus cluster, which is estimated to have a much stronger magnetic field compared to the GMF~\cite{2006MNRAS.368.1500T}. We will thus work with this source as an illustrative example, noting that an analysis combining multiple sources with careful modeling of their line-of-sight magnetic fields could potentially give improved sensitivity.

NGC1275 has a Galactic longitude of $150.58^{\circ}$ and a Galactic latitude of $-13.26^{\circ}$; its estimated distance is 68.2 Mpc. The emission spectrum of NGC1275 is also modeled by a log-normal representation (as in Eq.~\ref{eq13}) with the best fit parameters \cite{Fermi-LAT:2019yla} again given in Table~\ref{table3}. We follow Ref.~\cite{Day:2018ckv} for the modeling of the magnetic field. Specifically, we take a central magnetic field strength of $25\mu$G as indicated in Ref.~\cite{2006MNRAS.368.1500T}. The magnetic field falls off radially as $B \propto n_e^{0.7}$ as modeled in Ref.~\cite{2009RMxAC..36..303B} for galaxy clusters. The radial behavior of electron density is modeled in Ref.~\cite{2003ApJ...590..225C} as
 \begin{linenomath}
\begin{equation}
  n_e=\frac{3.9\times 10^{-2}}{(1+(r/80))^{1.8}}+\frac{4.05 \times 10^{-3}}{(1+(r/280)^2)^{0.87}}\text{cm}^{-3}. 
\end{equation}
\end{linenomath} 

There is not much known on the coherence length of the B-field of the Perseus cluster, so we parametrize it (as in Ref.~\cite{Day:2018ckv}) using the values motivated by a study of the structure of the magnetic field in the cool core cluster A2199~\cite{article}. The minimum coherence length is assumed to be 3.5kpc and the maximum coherence length 10kpc. The magnetic field along the line of sight is simulated with the above setup; a set of coherence lengths is drawn randomly from the probability distribution $P(l=x) \propto x^{-1.2}$, until the sum of the lengths adds up to 1 Mpc. In each domain, the magnetic field is assumed to be constant with fixed random direction.

As discussed previously, the radiation from the AGN could have a non-negligible polarized component due to synchrotron radiation, but for simplicity we will assume an initially unpolarized source for this analysis. 

 \subsection{\label{examples}Expected perturbation of the gamma-ray polarization}

\begin{figure*}[htbp]
\begin{center}
  \includegraphics[width=1\linewidth]{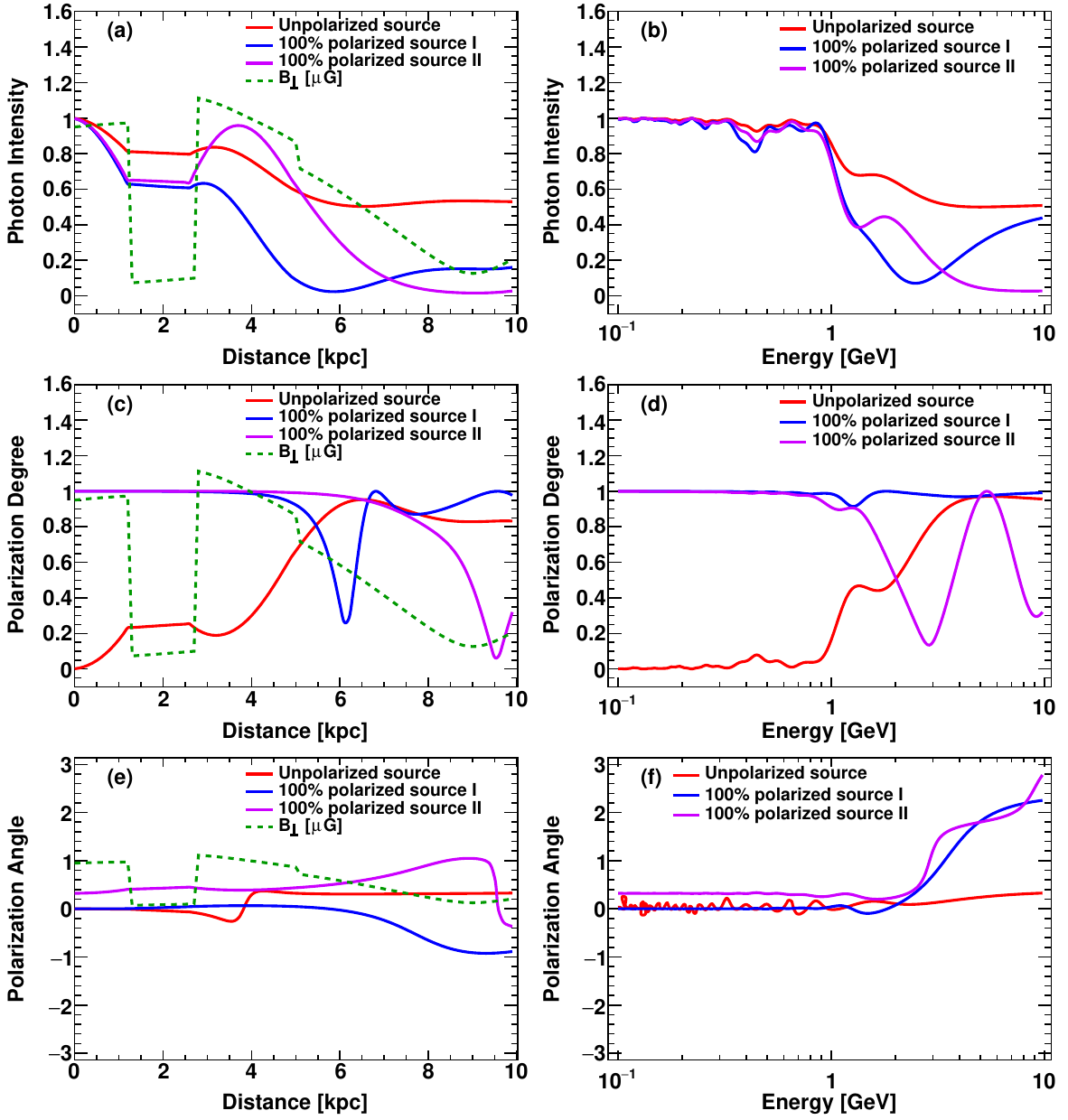}
 \caption{Example of expected photon intensity and polarization from the pulsar PSR J2021+3651, with ALP parameters $g_{a\gamma\gamma}=3.543\times 10^{-10}$ GeV$^{-1}$, $m_a = 4.41 \times 10^{-9}$ eV. (a) The intensity of photons at 3~GeV versus the distance along the line of sight from the pulsar.  (b) The intensity of photons versus the photon energy. (c) The polarization degree of photons at 3~GeV versus the distance along the line of sight from the pulsar. 
 (d) The polarization degree of photons versus the photon energy. (e) The polarization angle of photons at 3~GeV versus the distance along the line of sight from the pulsar. (f) The polarization angle of photons versus the photon energy, at the Earth's location. The red curve assumes an initially unpolarized source; the blue and violet curves assume initially 100\% polarized source with different  initial angles (see text for details). The transversal magnetic field $\mathrm{B_\perp}$ in units of $\mu$G is shown as the dashed green line in (a), (c), and (e). 
}
  \label{fig2}
 \end{center}
 \end{figure*}

By solving Eq.~\ref{eq2} numerically, we obtain the intensity, polarization degree, and polarization angle of photons as functions of both the distance along the line of sight from the pulsars and the photon energy. As an example,  Fig.~\ref{fig2} shows the calculated results for PSR J2021+3651, approximately matching the best-fit parameters for the ALP explanation of the excess claimed in Ref.~\cite{Majumdar:2018sbv}: specifically, we take $g_{a\gamma\gamma}=3.543\times 10^{-10}$ GeV$^{-1}$ and $m_a = 4.41 \times 10^{-9}$ eV. The photon-ALP mixing is related to the strength of the transversal magnetic field as shown in Fig.~\ref{fig2}(a), (c), and (e). This effect will induce modification of gamma ray spectra as shown in Fig.~\ref{fig2}(b). The intensity results in Figs.~\ref{fig2}(a-b) are  similar to the calculation in Ref.~\cite{Majumdar:2018sbv}, which we employ as a cross-check, although we use an updated magnetic field model so the results are not identical. 

As we can see from Eq.~\ref{eq3}, only the component parallel to the magnetic field will oscillate, so the final spectrum depends on the initial polarization state and the initial polarization angle. Fig.~\ref{fig2}(d) shows the polarization degree of photons as a function of photon energy, assuming three cases for the initial polarization: an initially unpolarized beam, and two initially 100\% linearly polarized sources with different initial polarization angles. The initial polarization of the blue curve aligns with the transverse external $B$-field at the position of the source, while the initial polarization angle of the purple curve is chosen such that the oscillation amplitude is maximized, which corresponds to $\tan \theta\approx 3$ (where $\theta$ is the angle between the polarization vector and the external transverse B-field at the source position). We observe that in this example, a large energy dependence can be obtained in the polarization degree, for either a polarized or unpolarized source (although for a polarized source it depends on the initial polarization angle). 
    
\section{\label{AMS}Polarization measurement by pair production}

\begin{figure}[htbp]
\begin{center}
  \includegraphics[width=1.0\linewidth]{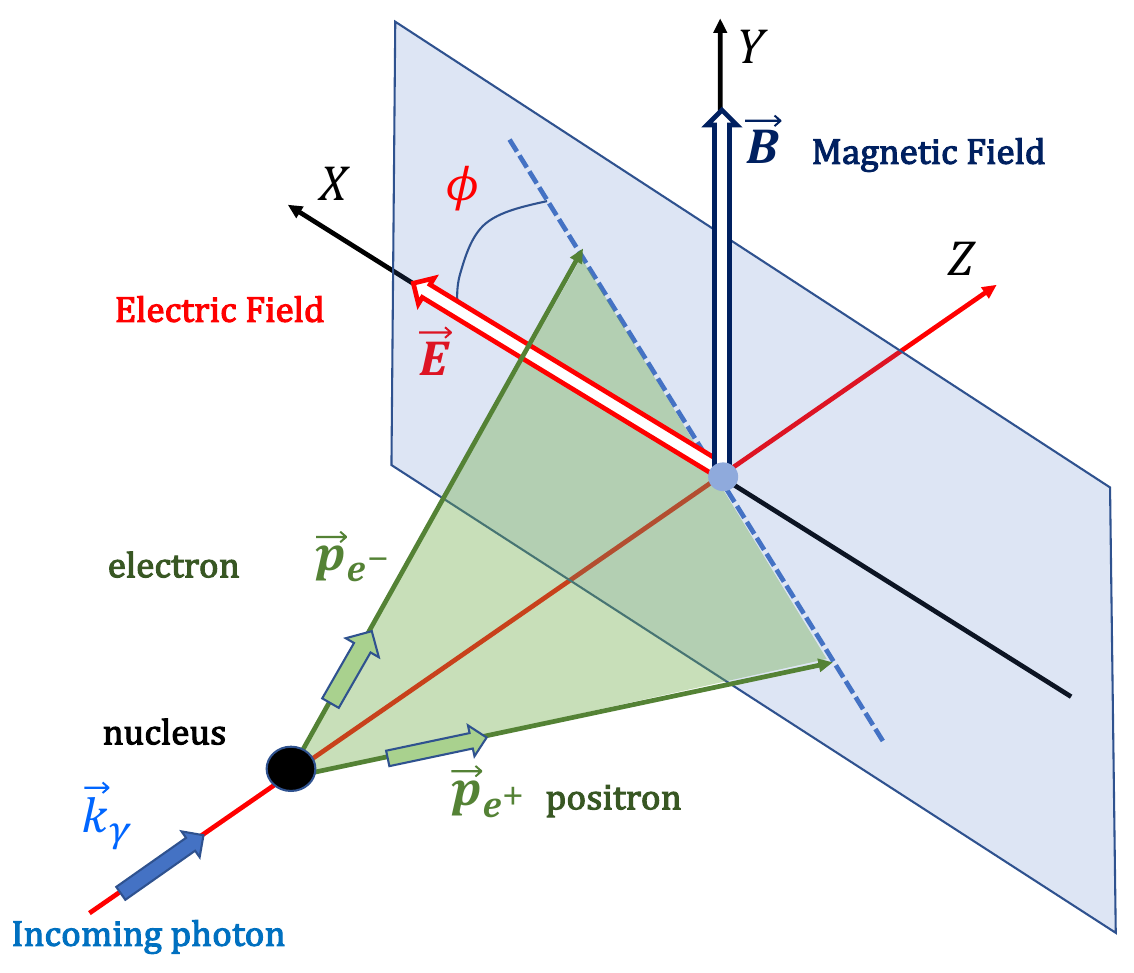}
 \caption{Kinematics of electron-positron pair production. The $x$ and $y$ axes are aligned with the directions of the electric field and magnetic field carried by the light source, respectively. The 3-momenta of the incoming photon, the emitted electron and positron are denoted as $\vec{k}_{\gamma}$,  $\vec{p}_{e^{-}}$, and $\vec{p}_{e^{+}}$. The $x-$ or electric field direction is also called the polarization direction of the source. The electron and positron form a plane (green) which is perpendicular to the $x-y$ plane (blue). The azimuthal angle $\phi$ is defined as the angle between the electron-positron plane and the direction of the electric field.}
  \label{fig:kinematics}
 \end{center}
 \end{figure}

Above the energy threshold of pair production ($\sim$1~MeV), the polarization of photons can be measured by the converted electron-positron pair. The detection of linearly polarized photons by pair production was proposed in the 1950s by Ref.~\cite{PhysRev.77.722.2,PhysRev.78.623}. The main signature from a polarized source of gamma rays is the preferential emitting direction of the electron-positron pair. The kinematics of electron-positron pair production is shown in Fig.~\ref{fig:kinematics}.  The azimuthal angle $\phi$ is defined as the angle between the electron-positron plane and the direction of the electric field of the source. The asymmetry of the azimuthal distribution carries information on the polarization of the source. 

Specifically, the photon event distribution over the azimuthal angle $\phi$ has the following form~\cite{Bernard:2013jea}:
\begin{linenomath}
\begin{equation}\label{eq4}
dN/d\phi \propto 1+A \cdot P \cos(2\phi)\ ,
\end{equation}
\end{linenomath}
where $A$ is the analyzing power depending on the angular resolution of the detector and the kinematics of the pair production, and $P$ is the degree of the photon polarization. For silicon-strip pair-production detectors operating in the energy range above 0.1 GeV, such as AMS-02, the value of $A$  can be approximated as a constant value 0.14~\cite{Gros:2016dmp, Bernard:2022jrj}. We will assume future successor instruments will have a similar value of $A$ and will differ only in their acceptance. That said, we caution that the effective polarization asymmetry can depend sensitively on the design of the detector; Ref.~\cite{Bernard:2022jrj} finds that the effective value of $A$ is significantly suppressed for the {\it Fermi}-LAT, for example. While that work argues that the AMS-02 tracker has a number of properties that are favorable for gamma-ray polarimetry (specifically, ``thin wafers, narrow readout pitch, and large distance between layers''), these properties would not necessarily be preserved in AMS-100. We explore the potential of AMS-100 for gamma-ray polarimetry with this assumption ($A\simeq 0.14$) in part to understand the science case for taking polarimetry into account in the design of AMS-100.

\begin{figure}[htbp]
\begin{center}
  \includegraphics[width=1\linewidth]{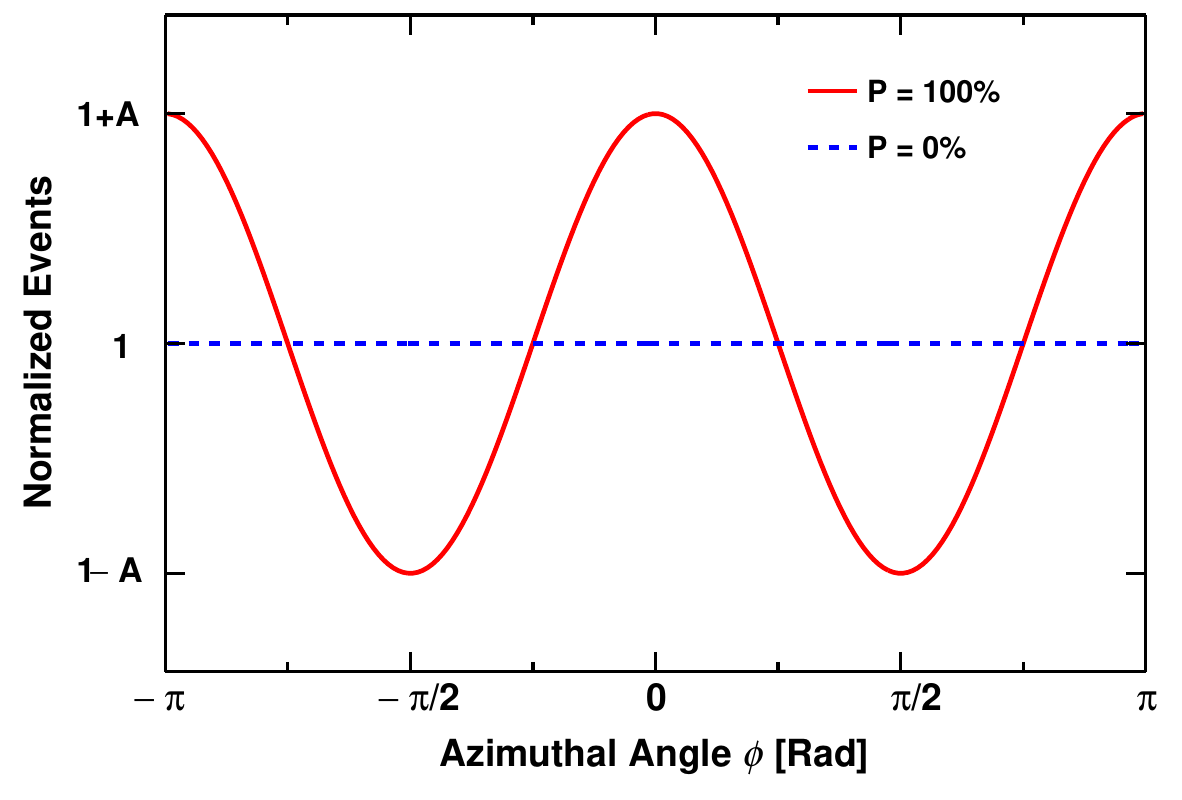}
 \caption{The expected event distribution as a function of azimuthal angle $\phi$ with fully polarized (red solid line) and unpolarized (blue dashed line) photons. A sinusoid shape is expected for polarized photons while a flat distribution is expected for unpolarized photons.}
  \label{fig:azimuthal}
 \end{center}
 \end{figure}

%a specific detector configuration 
Fig.~\ref{fig:azimuthal} shows the expected observation in azimuthal distribution with fully polarized ($P=100\%$) and unpolarized ($P=0\%$) photons; a sinusoidal shape is expected for polarized photons while a flat distribution is expected for unpolarized photons. The sinusoidal shape of the distribution indicates that for polarized photons, the electron-positron plane is preferentially aligned with the electric field of the source. The degree of polarization $P$ is determined by the magnitude of the sinusoidal oscillation.

The measurement of gamma rays with the AMS-02 experiment is described in Ref.~\cite{Beischer:2020rts}. The effective acceptance using photons converted in the upper detector increases as the energy increases from nearly zero at $\sim100$~MeV, reaching 140~cm$^2$sr at $\sim$ 5~GeV, and then decreases to zero at 1~TeV. For photon sources close to the zenith of AMS-02, the effective area is $\sim$ 180~cm$^2$. For point sources, it is the exposure  rather than the overall acceptance that is important. A realistic calculation of the AMS-02 exposure over six years, accounting for the efficiency of various selection cuts, is conducted in Ref.~\cite{Beischer:2020rts} for one specific energy (2 GeV). For our AMS-02 calculations, we digitize the results of that work for the exposure (for the conversion analysis, not the calorimeter events) at the positions of our sources. We then rescale the results by a factor of $10/6$ to obtain a 10-year sensitivity. The resulting exposures (at 2 GeV) are described in Table~\ref{table4}. We extend the results to our full energy range (100 MeV -- 10 GeV) by assuming the energy dependence of the exposure is dominated by the energy dependence of the effective area, which is given for the conversion analysis in that same work.

 \begin{table}[htbp]
  \caption{Estimated signal and background (bkg) counts for AMS-02 in ten years for selected sources, along with exposure at 2 GeV.  The first six entries correspond to pulsars, the next four to supernovae, and the last three to our example extragalactic source and Geminga and Vela pulsars.}
  \begin{tabular}{|m{0.2\linewidth}|m{0.2\linewidth}|m{0.2\linewidth}|m{0.2\linewidth}|}
  \hline
   \centering Source name & \centering Estimated counts for AMS-02 in ten years & \centering Estimated bkg counts for AMS-02 in ten years & \centering Exposure at 2 GeV \quad [$10^9$ cm$^2$
   s]\tabularnewline
  \hline
  \centering J1420-6048 & \centering 14 & \centering 62& \centering 0.308\tabularnewline
   \centering J1648-4611 & \centering 6 & \centering 194& \centering 0.729\tabularnewline
     \centering J1702-4128 & \centering 8 & \centering 250& \centering 0.866\tabularnewline
     \centering J1718-3825 & \centering 29 & \centering 200& \centering 0.923\tabularnewline
  \centering J2021+3651 & \centering 246 & \centering 173& \centering 1.79\tabularnewline   
   \centering J2240+5832 & \centering 13 & \centering 71& \centering 1.70\tabularnewline
   \centering IC443 & \centering 185 & \centering 49& \centering 1.60\tabularnewline   
    \centering W44 & \centering 209 & \centering 357& \centering 1.43\tabularnewline   
    \centering W51C & \centering 124 & \centering 219& \centering 1.52\tabularnewline     
    \centering W49B & \centering 53 & \centering 272& \centering 1.48\tabularnewline  
    \centering NGC1275 & \centering 133 & \centering 13& \centering 1.82\tabularnewline  
    \centering Geminga & \centering 2244 & \centering 13& \centering 1.62\tabularnewline  
    \centering Vela & \centering 2191 & \centering 12& \centering 0.737\tabularnewline  
     \hline    
  \end{tabular}
  \label{table4}
  \end{table}

The proposed next-generation magnetic spectrometer AMS-100 aims for an effective acceptance of 30~m$^2$sr for photon conversions in the silicon tracker layers \cite{Schael:2019lvx}. Compared to the peak effective acceptance for AMS-02 conversion events, which is 140~cm$^2$sr as discussed above, we expect an increase in statistics by roughly a factor of 2000, averaged over the whole sky.

Since AMS-100 would be located at Lagrange Point 2 and has a very different (cylindrical) geometry to AMS-02, rather than rescale the AMS-02 exposure, we simply assume a uniform exposure over the sky for AMS-100. This leads to an average exposure over 10 years of $30$ m$^2$sr $\times$ 10 yr$/(4\pi \, \text{sr})= 7.5 \times 10^{12} \text{cm}^2$ s. We furthermore assume that this corresponds to the peak acceptance/exposure as a function of energy, and rescale the exposure at lower and higher energies according to the energy-dependence of the AMS-02 effective area, since the silicon tracker design is similar.

Ref.~\cite{Schael:2019lvx} also suggests a smaller AMS-100 pathfinder. The pathfinder would have geometrical acceptance of $0.2\times$ the AMS-02 value, so we simply scale down the exposure by this factor. (Note that Ref.~\cite{Schael:2019lvx} characterizes this pathfinder as ``$10\%$ scale'', but roughly a factor of 2 of this loss would come from removing the calorimeter, which is irrelevant for our analysis.) We hereafter denote this $20\%$-scale pathfinder as AMS-100P.

Having obtained the energy-dependent exposure for both telescopes, we integrate these functions over the energy spectra of the sources (as given in Tables \ref{table2}-\ref{table3}) to obtain the total number of expected counts.

%The proposed 
%next generation magnetic spectrometer AMS-100 will increase the acceptance of gamma rays by at least a factor of $\sim$ 100~\cite{Schael:2019lvx}.

\section{\label{sensitivity}Estimating sensitivity}
%To simulate the observation by AMS,

Having introduced the physics behind detection, in this section, we will make projections for both AMS-02 and future AMS-like detectors. If the gamma-ray polarization of the sources was well-understood, we could perform a (forecast) likelihood analysis for the detectability of a signal from new physics; however, as discussed above, we do not have such a model for the source polarization. We will thus perform two more generic analyses in this section, addressing sensitivity to (1) a non-zero polarized fraction from an initially unpolarized source (unbinned analysis) and (2) variation in the polarization fraction as a function of energy  (binned analysis). The first analysis is appropriate for a first measurement of GeV-energy gamma-ray polarization from a given source, and for constraining the effects of ALPs via the polarization they should inevitably induce in gamma-rays from initially unpolarized sources. In the event of a detected polarization signal, the second analysis would potentially allow us to distinguish ALP-induced polarization from other sources of polarization, which would not be expected to have the same energy dependence. A null result in either channel could be used to set upper limits on an ALP-induced polarization signal; our analysis will focus on detection sensitivity, but we will discuss the implications of a null result in section \ref{sec:results}.

   \subsection{\label{genericpolsearch}Sensitivity to polarization fraction}

We first investigate the detectability of a non-zero polarization for AMS-like detectors, without specifying the source of the polarization. This analysis is broadly applicable and can be employed to estimate the ability of these detectors to measure polarization induced by ALP-photon mixing from an initially unpolarized source.

An analytical expression for the minimum detectable polarization (MDP) can be given in terms of the observed counts by~\cite{Giomi:2016brf}: 

 \begin{linenomath}
\begin{equation}
  \text{MDP}(p)=\frac{2\sqrt{-\ln(p)}}{A}\frac{\sqrt{N_S+N_B}}{N_S}, \label{eq:MDP}
\end{equation}
\end{linenomath}
where $A=0.14$ for AMS-02 as defined before, $N_S$ and $N_B$ are signal and background counts respectively, and $p$ is the probability threshold for a detection, i.e.~above the MDP($p$) polarization threshold, the associated modulation amplitude has probability $\le p$ that it would be exceeded by a statistical fluctuation if the true polarization fraction were zero. Note that while the MDP is commonly used as an estimate of the sensitivity of a polarimeter, it is not the same as (for example) ``the true polarization value for which the zero-polarization hypothesis is expected to be excluded with a given $p$-value'' (see also \cite{2010SPIE.7732E..0EW} for discussion); the latter quantity depends on the uncertainty in the extraction of the polarization amplitude when the true polarization is non-zero, whereas MDP is calculated purely under the zero-polarization hypothesis. However, it is useful as an estimate for the number of counts needed to render polarization potentially detectable.

As a first optimistic estimate, we take $N_B=0$.  We see that for the MDP to be below 1 with $p=0.05$, we need $N_S > (2\sqrt{-\ln0.05}/0.14)^2 \approx 600$. This sets a floor for the brightness of sources where we can hope to measure polarization.

For each of the sources we consider, we compute the expected number of photons in 10 years of AMS-02 or AMS-100 data, estimating the exposure as described in the previous section.
%in 10 years of AMS-02 data. We estimate the total detected events by integrating the emission spectra multiplied by the acceptance of the AMS-02 detector (described in Ref.~\cite{Beischer:2020rts}) over a period of ten years (corresponding to the operation time of the AMS-02 detector). As for the AMS-100 detector, we assume its acceptance is 100 times of that of AMS-02.
In Table~\ref{table4} we show the estimated counts for each source in AMS-02. We see that of the sources we consider, only the Geminga and Vela pulsars are expected to have enough photons to be above the (zero-background) threshold at which MDP $< 1$. Ignoring the background counts (which as we will show are quite small for both Geminga and Vela), the MDP is 0.52 for Geminga, and 0.53 for Vela.

%\hl{we show the resulting MDP for each of our selected sources using AMS-02. We should note that for some of the sources,  AMS-02 will not have enough sensitivity to distinguish an unpolarized state even from a fully polarized state. We see that for both the highest-statistics pulsars and supernova remnants, the MDP is around $30\%$; i.e. a polarization fraction above $\sim 30\%$ would correspond to an amplitude that has less than $5\%$ probability if the emission is unpolarized. For example, for the supernova remnant W44 we expect 5542.31 events in 10 years of AMS-02 data, corresponding to a MDP of 0.33 at 95\% confidence.}

What of the backgrounds? A full analysis including the diffuse photon background and the background from misidentified cosmic rays would take into account the probability that each event is associated with the source vs being part of the background, based on its distance from the center of the source and the point spread function of the instrument.

Such an in-depth analysis is beyond the scope of this article; instead, we estimate the background by computing the  background within a $1^\circ$ radius of each source, which corresponds roughly to the 68\% containment region of the AMS-02 point spread function at an energy of 700 MeV \cite{Beischer:2020rts}. The angular resolution improves at higher energies, to slightly below 0.1$^\circ$ (in 68\% containment angle) at 10 GeV, and degrades at lower energies, to around $3^\circ$ at 200 MeV (below this energy the effective area for conversion events becomes negligible). We could average this angular resolution function over energy, weighting by the expected photon counts; this inherits the energy dependence of the effective area, and we will also assume a photon source with a spectral index of $dN/dE \sim E^{-2}-E^{-3}$ (typical of the Galactic diffuse emission). For this range of photon spectra we obtain photon-number-averaged values for the 68\% containment angle of $\sim 0.9-1.5^\circ$, supporting our use of $1^\circ$ as an estimate.

Before we go on to discuss the modeling of the diffuse photon background, let us briefly discuss misidentified cosmic rays. Ref.~\cite{Beischer:2020rts}, which we use for our exposure estimates for AMS-02 (including selection cuts), suggests that in a broad region including the Galactic plane (Galactic latitude $|b|<8^\circ$) the cosmic-ray background is generally subdominant to the diffuse photon background. However, not all our sources are along the plane. Ref.~\cite{Beischer:2020rts} also provides an estimate for cosmic-ray background events based on the observed difference between the measured and modeled gamma ray signals; they show that the background is declination-dependent, but has a maximum value of around 20,000 events/sr at high declinations (at lower declinations this rate is suppressed by a factor of around two), in 6 years of AMS-02 data. This corresponds to 19 events in a $1^\circ$ radius circle in the 6-year dataset of that work, so roughly 32 events over 10 years. Thus, for any source where the minimum detectable polarization is below 1 (requiring $N_S > 600$), we see that this background must be negligible ($\le 5\%$) compared to the signal, with the ratio $\sqrt{N_S + N_B}/N_S$ thus changing only at the percent level due to the inclusion of this background. For this reason, we will ignore the cosmic-ray background for the remainder of our analysis.

In contrast, the diffuse photon background can be important, especially for relatively faint sources lying in the Galactic plane. We employ the Pass 8 (P8R3) model for the diffuse Galactic gamma-ray emission provided by the {\it Fermi}-LAT Collaboration\footnote{\url{https://fermi.gsfc.nasa.gov/ssc/data/access/lat/BackgroundModels.html}} to obtain the photon flux at the location of each source (as a function of energy), and then use the previously-derived exposure at the source location to convert this flux to a number of counts. The results are shown in Table~\ref{table4}, for the total counts integrated from 0.1 GeV to 10 GeV. We see that for sources in the Galactic plane, the backgrounds are often comparable to the counts from the sources themselves; however, for NGC1275 and the Geminga and Vela pulsars, which are further from the Galactic plane, the backgrounds are expected to be quite subdominant. 

As a result, for the sources where polarization may be detectable with AMS-02 -- i.e. the Geminga and Vela pulsars -- we expect the zero-background MDP estimate to be quite accurate.

For the other sources, we will now move on to the prospects with AMS-100. Table \ref{table5} shows the estimated MDP both with and without background events. As we see, for the best sources in all categories (Galactic pulsars, Galactic supernovae, and the extragalactic source NGC1275), we expect to have MDP at the 1-4\% level even in the presence of backgrounds, and the backgrounds do not markedly degrade the estimated MDP, although for other sources the impact of backgrounds can be substantial.   

Finally, with these diffuse photon background results in hand, let us briefly revisit our neglect of the cosmic-ray background. We see that for Vela, Geminga, and NGC1275, the two backgrounds are of the same order; however, they are both very subdominant compared to the source brightness. For all other sources, the diffuse photon background dominates over the cosmic-ray background. Thus we expect our neglect of the cosmic-ray background to have no qualitative effect on our results. If AMS-100 sought to study faint sources far from the Galactic plane in polarization, the cosmic-ray background might then become a limiting factor.

  \begin{table}
  \caption{Estimated signal and background counts for AMS-100 in ten years with minimal detectable polarization (MDP) for selected sources.   The first six entries correspond to pulsars, the next four to supernovae, and the last three to our example extragalactic source and Geminga and Vela pulsars. 1 should be understood as there is no way to distinguish a polarized source from an unpolarized one with the statistics.}
  \begin{tabular}{|m{0.2\linewidth}|m{0.2\linewidth}|m{0.2\linewidth}|m{0.15\linewidth}|m{0.15\linewidth}|}
  \hline
   \centering Source name & \centering Estimated counts for AMS-100 in ten years & \centering Estimated bkg counts for AMS-100 in ten years & \centering MDP for AMS-100& \centering MDP for AMS-100 with bkg\tabularnewline
  \hline
  \centering J1420-6048 & \centering 291744& \centering 1287312 & \centering 0.05 & \centering 0.11\tabularnewline
   \centering J1648-4611 & \centering 53938& \centering 1704463 & \centering 0.11 & \centering 0.61\tabularnewline
     \centering J1702-4128 & \centering 62565& \centering 1847690 & \centering 0.10 & \centering 0.55 \tabularnewline
     \centering J1718-3825 & \centering 198806& \centering 1386797 & \centering 0.06 & \centering 0.16 \tabularnewline
  \centering J2021+3651 & \centering 881567& \centering 619863 & \centering0.03 & \centering0.03 \tabularnewline   
   \centering J2240+5832 & \centering 49832& \centering 269155 & \centering 0.11 & \centering 0.28 \tabularnewline
   \centering IC443 & \centering 741047& \centering  196893 & \centering 0.03 & \centering 0.03\tabularnewline   
    \centering W44 & \centering 938385& \centering  1603193 & \centering  0.03  & \centering 0.04 \tabularnewline   
    \centering W51C & \centering 522037& \centering  926101 & \centering  0.03  & \centering 0.06\tabularnewline     
    \centering W49B & \centering 230353& \centering  1174738 & \centering  0.05  & \centering 0.13\tabularnewline  
    \centering NGC1275 & \centering 468941& \centering  44404 & \centering  0.04 & \centering 0.04\tabularnewline  
    \centering Geminga & \centering 8853197& \centering  49339 & \centering  0.01  & \centering 0.01\tabularnewline  
    \centering Vela & \centering 19042725& \centering  102915 & \centering  0.01  & \centering 0.01\tabularnewline  
     \hline    
  \end{tabular}
  \label{table5}
  \end{table}

\subsection{Monte Carlo simulations for ALP parameter space}

In this light, we now move on to study the degree to which AMS-100 could constrain ALP parameter space. (We do not perform this analysis for AMS-02 given the results of the simple MDP estimate, which suggest that none of the more distant sources will have the statistics required to measure any level of polarization.)

For each of the sources discussed in this work, and for each point $\{m_a,g_{a\gamma \gamma}\}$ in ALP parameter space, we calculate the expected photon polarization spectrum at the AMS-100 detector using the modeling described in section \ref{sec:Pheno}. We compute the expected counts and their polarization fraction over the full 0.1-10 GeV energy range, and in four or ten log-spaced energy bins, using the detector properties described in section \ref{AMS}. For each choice of binning, we average the forecast linear polarization fractions as a scalar sum  over the total counts within each bin (weighted by the number of photons as a function of energy) and use it as the true polarization fraction for the Monte Carlo (MC) simulations. To treat it more rigorously, one should average the individual polarization+intensity vectors, using a vector sum over counts instead. In practice, we have confirmed that since the polarization angle varies continuously without sudden jumps, the two averaging methods yield similar results.  The spectral distortion due to the bin-to-bin migration is negligible for these wide bins. 

We then simulate the observed counts as a function of azimuthal angle (independently by energy bin where appropriate), treating the counts as random Poisson variables in 10 uniform angular bins from 0 to $2\pi$.  The mean value for each Poisson distribution is predicted by Eq.~\ref{eq4} using the true polarization as input for the polarization degree $P$, and determining the normalization from the expected detected events with $A=0.14$. We perform 1000 draws from the Poisson distribution to account for the randomness. 

For NGC1275 there is an additional subtlety that the magnetic field involves a random draw. We draw 100 different $B$-field configurations for NGC1275, and for each ALP parameter point, we simulate 1000 realizations for each $B$-field configuration.

For each realization (and each energy bin where appropriate), the event distribution as a function of azimuthal angle $\phi$, with the error approximated by the square root of the number of counts,\footnote{This approximation should be valid in the limit of a large number of counts; as discussed above, where the number of counts is $\ll 600$ we will not be able to detect even a large polarization fraction, so this should be a good approximation for the parameter space of interest.} is fitted to the functional form of Eq.~\ref{eq4} to obtain the reconstructed polarization degree and its uncertainty (using the Scipy \texttt{curve\_fit} function, which implements a weighted least-squares fit). We confirmed that the distribution of reconstructed polarization is consistent with the input to the simulation, and that the error bars are consistent with the scatter across simulations. Thus for each realization (mock dataset) and for each source, we have an estimate for the polarization in the $i$th energy bin as $P_i \pm \Delta P_i$. These are the data that enter into our subsequent analyses.

\begin{figure}[htbp]
\begin{center}
  \includegraphics[width=0.95\linewidth]{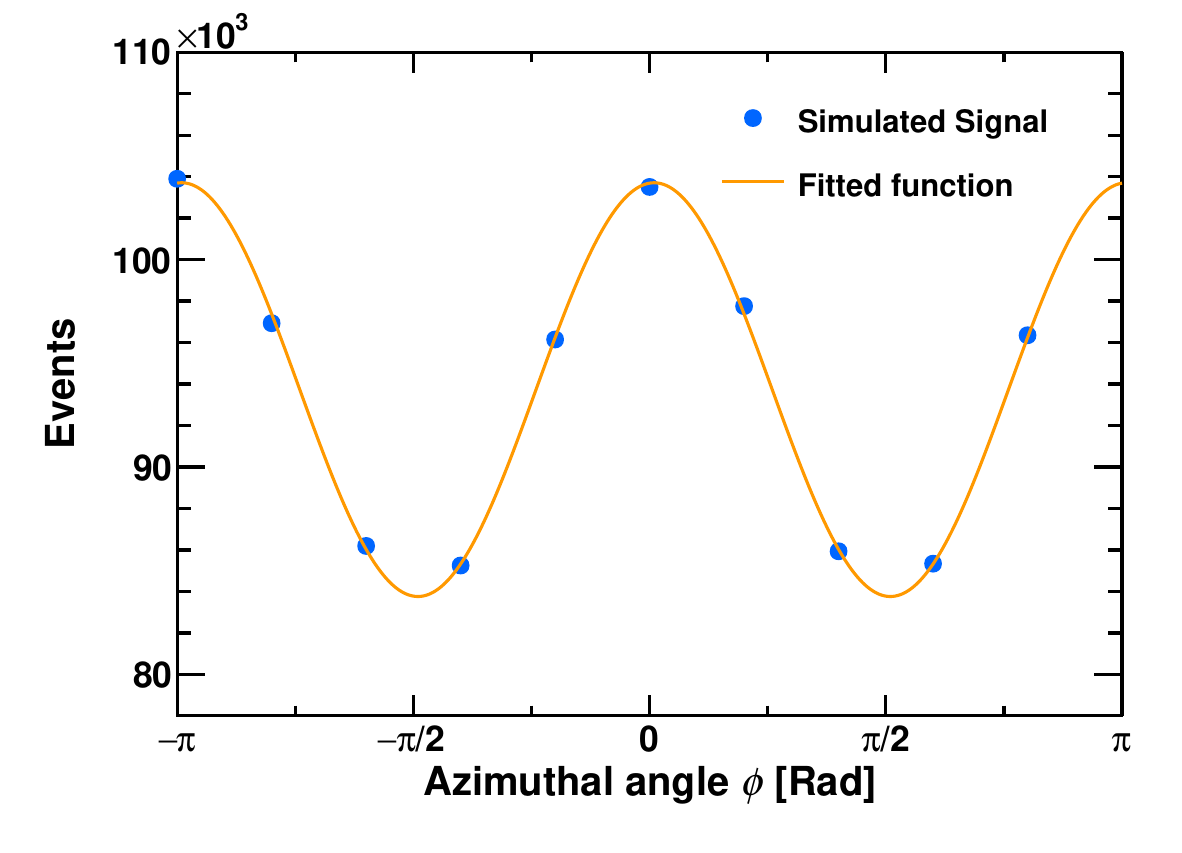}
 \caption{The event distribution as a function of azimuthal angle $\phi$ with initially unpolarized photons from W44 in the unbinned analysis, in one example realization. We assume ten years of exposure with AMS-100. The orange line is the fitted function and the blue dots are results from MC simulation. The ALP parameters are $g_{a\gamma \gamma} = 3.543\times 10^{-10}$ GeV$^{-1}$, $m_a = 4.41\times 10^{-9}$ eV.}
  \label{fig:fit1}
 \end{center}
 \end{figure}

 Fig.~\ref{fig:fit1} shows an example of a simulated event distribution (in azimuthal angle) for a unbinned (in energy) analysis for the W44 supernova remnant, assuming initially unpolarized photons with ALP parameters $g_{a\gamma\gamma}=3.543\times 10^{-10}$ GeV$^{-1}$ and $m_a = 4.41 \times 10^{-9}$ eV. As previously, W44 gives rise to 938385 expected counts over ten years with AMS-100.
 
 \subsection{Statistical tests for polarization and energy dependence}

 Given a set of polarization values and error bars for each energy bin, we are interested in determining:
 \begin{itemize}
     \item In the case of a single large energy bin, the $p$-value of the (mock) data with respect to the zero-polarization hypothesis.
     \item In the case of multiple energy bins, the $p$-value of the (mock) data with respect to the hypothesis of energy-independent polarization.
 \end{itemize}
 
To compute the $p$-values we employ a $\chi^2$ test. For Galactic sources, having determined these results for each realization, we average the $p$-value across the 1000 realizations to obtain the expected $p$-value (for exclusion of the null hypothesis / detection of a signal).

For NGC1275 analyses, we need to take into account the uncertainties in the $B$-field model. For each ALP parameter point and each of the 100 choices of $B$-field configuration, we evaluate the expected $p$-value as described above (averaging over 1000 realizations), and then rank the $B$-field configurations by the expected $p$-value (independently for each ALP parameter point). We will show results for the median $B$-field configuration by this metric (i.e. $50\%$ of B-field configurations would predict a higher expected $p$-value and $50\%$ lower), as well as for an ``unfavorable'' $B$-field configuration at the $95$th percentile of $p$-value (i.e. $95\%$ of B-field configurations would predict a lower $p$-value / higher significance of detection).  

\subsubsection{Detecting ALP-induced polarization from an initially unpolarized source}
\label{nonxeropolar}

The first test we perform examines the detectability of a non-zero polarization degree generated by ALP-photon mixing; the null hypothesis is that the signal is unpolarized.  We use one large energy bin, so the data for each realization are a single polarization measurement $P\pm \Delta P$, and the null hypothesis (zero polarization) has no free parameters, so there is only one degree of freedom per source. We perform separate analyses for the combined Galactic supernova remnants on one hand, and the extragalactic source NGC1275 on the other hand. We evaluate the test statistic (TS), for a combination of sources, as:
\begin{equation}
\chi^2 = \sum_\text{sources} \left(\frac{P}{\Delta P}\right)^2, \label{eq:chisq1}
\end{equation}
and convert this result to a $p$-value using the $\chi^2$ distribution with a number of degrees of freedom equal to the number of sources. 

An alternative TS would be to compute the Poisson likelihood of drawing the observed distribution of counts with respect to azimuthal angle, under (1) the zero-polarization hypothesis and (2) the alternative hypothesis where the polarization fraction is allowed to vary. Maximizing the likelihood ratio $\mathcal{L}/\mathcal{L}_0$ gives a best-fit value for the polarization fraction, and $2 \ln \mathcal{L}_\text{max}/\mathcal{L}_0$ then yields a TS which is $\chi^2$-distributed under certain assumptions. Eq.~\ref{eq:chisq1}  provides an approximation to this TS when the likelihood can be approximated as Gaussian.

The assumption of a $\chi^2$ distribution for the TS is expected to be imperfect, in particular because the null hypothesis here corresponds to a boundary of the parameter space. This can be addressed by simulating the distribution of our TS under the null (zero polarization) hypothesis.

\begin{figure}[htbp]
\begin{center}
  \includegraphics[width=1.\linewidth]{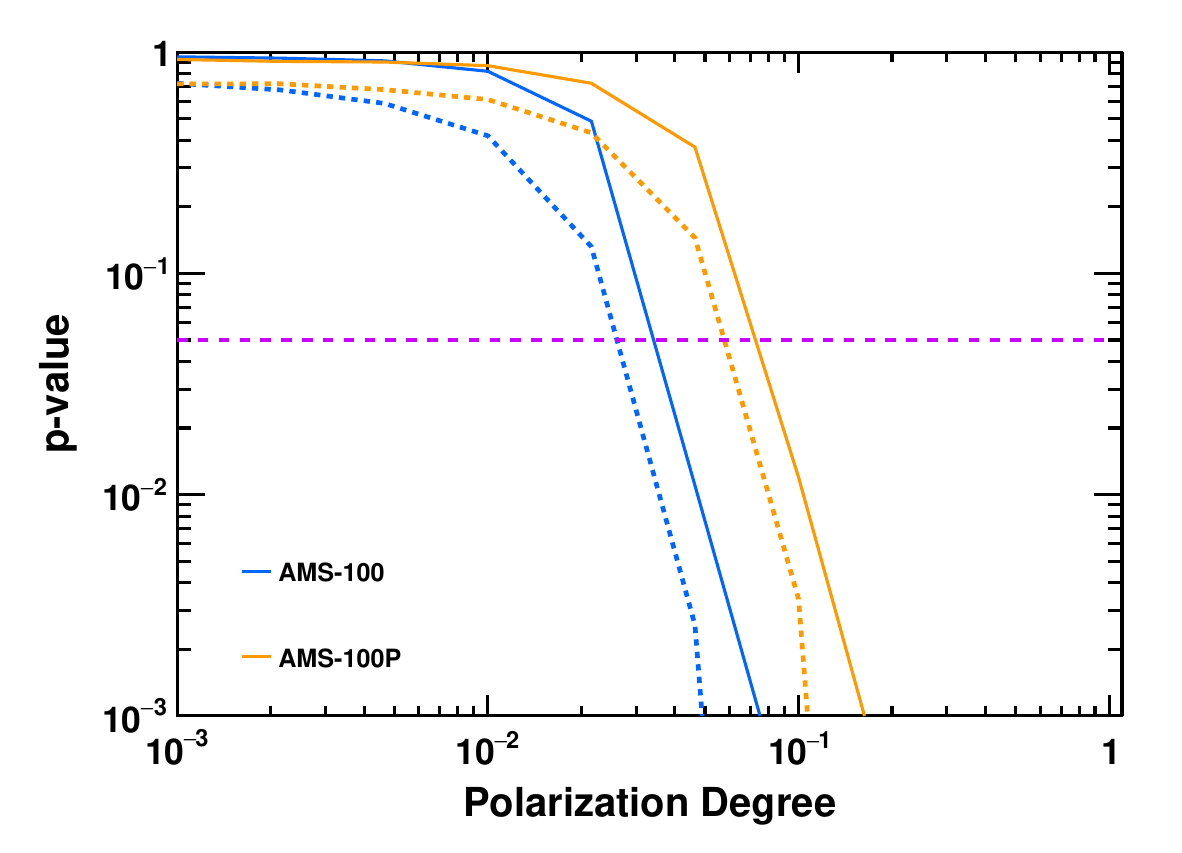}
 \caption{The expected $p$-value for exclusion of an unpolarized signal, as a function of the photon polarization on arrival of the signal at the detector, for a photon count rate consistent with the supernova W44 (i.e.~938385 counts in AMS-100 data as blue lines and $20\%$ of that as orange lines). The dotted line corresponds to our first approach, deriving TS from the log likelihood ratio and determining the $p$-value  using Monte Carlo. The solid line corresponds to the second (simplified) approach, where we make a Gaussian-likelihood approximation for the TS and approximate its distribution under the null hypothesis by a $\chi^2$ distribution. The dashed purple line corresponds to $p$-value=0.05.}
  \label{fig:pd}
 \end{center}
 \end{figure}

 As a check on our simulation pipeline and approximations, we fixed the polarization fraction of the incoming photons (rather than predicting it from the ALP parameters), and computed the relationship between the true polarization and the expected $p$-value for exclusion of the null hypothesis, in two ways:
 \begin{enumerate}
 \item Choosing the TS to be $2 \ln \mathcal{L}_\text{max}/\mathcal{L}_0$, we computed the distribution of this TS under the null (zero polarization) hypothesis. We then evaluated this TS in simulations with a fixed larger polarization fraction, and converted to a $p$-value based on the probability of finding an equal or higher TS in the null-hypothesis simulations.
 \item Choosing the TS as given in Eq.~\ref{eq:chisq1}, we evaluated this TS in simulations with a fixed larger polarization fraction, and converted to a $p$-value  using the $\chi^2$ distribution with an appropriate number of degrees of freedom (as described above).
 \end{enumerate}
 In Fig.~\ref{fig:pd} we show the result of these two calculations as a function of the true polarization, for an example source. We found that the two calculations are generally in reasonable agreement, and the second (simpler) approach always slightly underestimates the constraining power of the analysis. Accordingly, we use the second method for the rest of our analyses, noting that this may lead to slightly weaker sensitivity forecasts. We attribute the fairly good agreement to the fact that achieving an expected $p$-value of 0.05 requires both a large number of photon counts (meaning the Poisson distribution can be well-approximated by a Gaussian) and that the true polarization is well away from the boundary value of zero polarization. In this case we also find quite close agreement between the polarization degree corresponding to $p=0.05$ and the analytic MDP calculation for $p=0.05$ (Eq.~\ref{eq:MDP}).

It is already apparent from  Table~\ref{table4} that detection of an ALP-induced polarization signal with AMS-02 is implausible, since polarization is only expected to be detectable at all in Geminga and Vela (which are not favorable targets for an ALP-induced polarization signal). Furthermore, the relatively high MDP even for these sources means that we do not expect to be able to obtain meaningful constraints on the energy dependence of the polarization fraction with AMS-02, even if a polarization signal is detected.

%\hl{will be very challenging with AMS-02. The simplest version of such an analysis would involve splitting the dataset into two bins; if we chose these bins to have roughly equal numbers of photon counts, the MDP at 95\% confidence would rise to roughly 0.5 in each bin (from 0.3 in the full dataset). In the maximally optimistic case where one bin had zero polarization and the other had 100\%, we would expect to only be able to say (at 95\% confidence) that one bin had below 50\% polarization and the other had above 50\%, which is only a very marginal detection of variation. Considering realistic scenarios where the polarization varies smoothly with energy, even before specifying the ALP model, we see that AMS-02 is unlikely to have the statistical power to distinguish such scenarios.

 % In contrast, for AMS-100, there will be ample events for both of these analysis. As a result, we will only show projections of detection of a non-zero polarized fraction for the AMS-02 detector; whereas for the future AMS-100 detector we will include the results of both this analysis and a search for energy-dependence in the polarization signal.

  However, for AMS-100 (or AMS-100P), many sources will have sufficiently high statistics to measure both polarization and its energy dependence. As a result, from this point on we will focus solely on AMS-100/AMS-100P when forecasting sensitivity to ALP parameters.

  \subsubsection{\label{variation}Detecting  variation of polarization in energy spectrum}

 In the event of a detection of polarization (i.e.~exclusion of the zero-polarization null hypothesis), from an initially polarized or unpolarized source, we could take the analysis a step further and investigate if the detected polarization exhibited the oscillatory behavior (with respect to energy) generated by ALP-photon mixing. Such an analysis would require, as a minimum, having the sensitivity to exclude a constant, energy-independent polarization fraction. Thus in this section we take the null hypothesis to be that the polarization signal does not vary with energy. We employed our simulations of ALP signatures to ask whether a future detector, such as AMS-100, could successfully exclude this hypothesis in the presence of an ALP signal. 

\begin{figure}[htbp]
\begin{center}
  \includegraphics[width=1.\linewidth]{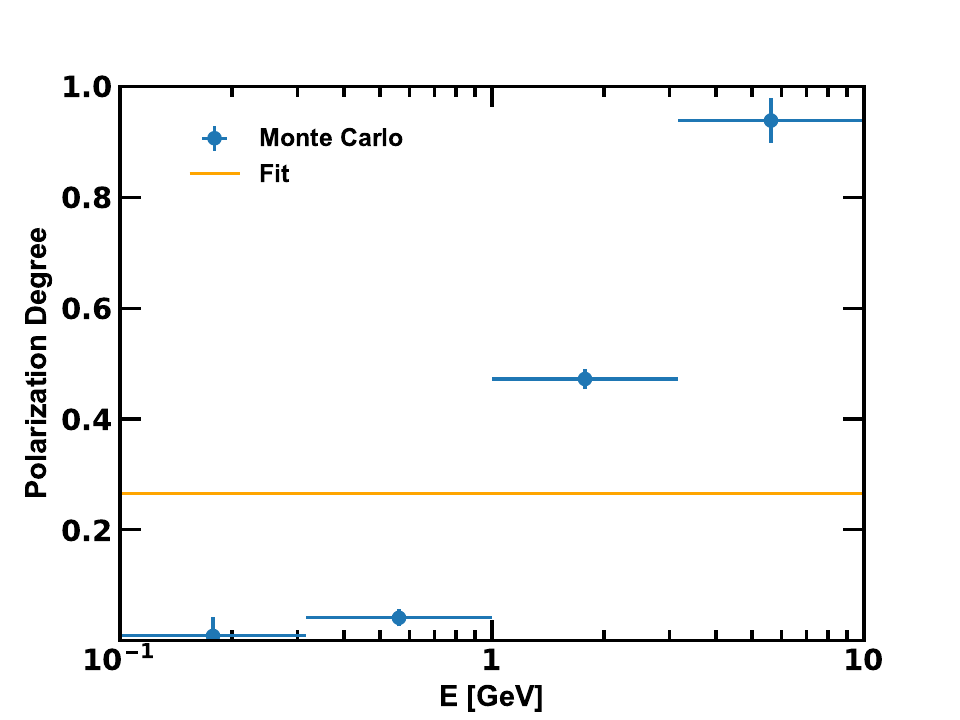}
 \caption{An example of fitting a constant line to an energy dependent polarization to test the variation for PSR J2021+3651. The blue dots are the expected polarizations in each energy bin and the orange line is the fitted constant linear polarization as the null hypothesis. The simulation assumes ten years of AMS-100 data, and the ALP parameters are $g_{a\gamma\gamma}=3.543\times 10^{-10}$ GeV$^{-1}$ and $m_a = 4.41 \times 10^{-9}$ eV.}
  \label{fig:fit2}
 \end{center}
 \end{figure}

We used a multi-bin $\chi^2$ test on our simulated data for the linear polarization spectra from 0.1 GeV to 10 GeV, with the energy range being divided equally in log scale into four (ten) bins for Galactic (extragalactic) sources (we discuss the dependence on the choice of binning in App.~\ref{app:binning}). We assume ten years of AMS-100 data. 
The simulations in each bin are similar to the one shown in Fig.~\ref{fig:fit1}, but with fewer counts. Given an inferred polarization fraction ($P_i$) and its error bar ($\Delta P_i$) in each bin, where $i$ indexes the energy bins, we minimize the $\chi^2$ for a one-parameter model with constant (energy-independent) polarization fraction, for each source independently, and then sum the $\chi^2$ over sources:
\begin{equation}\chi^2 = \sum_\text{sources} \sum_i \left(\frac{P_i - P_\text{model}}{\Delta P_i}\right)^2 \label{eq:chisq2} \end{equation}
We then translate the $\chi^2$ test statistic for this model to a $p$-value for exclusion, using the $\chi^2$ distribution with the appropriate number of degrees of freedom. Since the model has one degree of freedom (per source), the number of degrees of freedom per source equals the number of energy bins minus one. The simulated data and best-fit constant-polarization model for one example are shown in Fig.~\ref{fig:fit2}.

We perform separate analyses for (1) NGC1275, (2) Galactic supernova remnants, and (3) Galactic pulsars (as described in section \ref{sources}), considering different possibilities for the initial polarization of the pulsars. As in the single-bin analysis, for the NGC1275 analysis we repeat the calculation for 100 different realizations of the NGC1275 magnetic field.

 \section{Results}
 \label{sec:results}

 \subsection{Extragalactic sources}

   \begin{figure*}[htbp]
\begin{center}
  \includegraphics[width=0.49\linewidth]{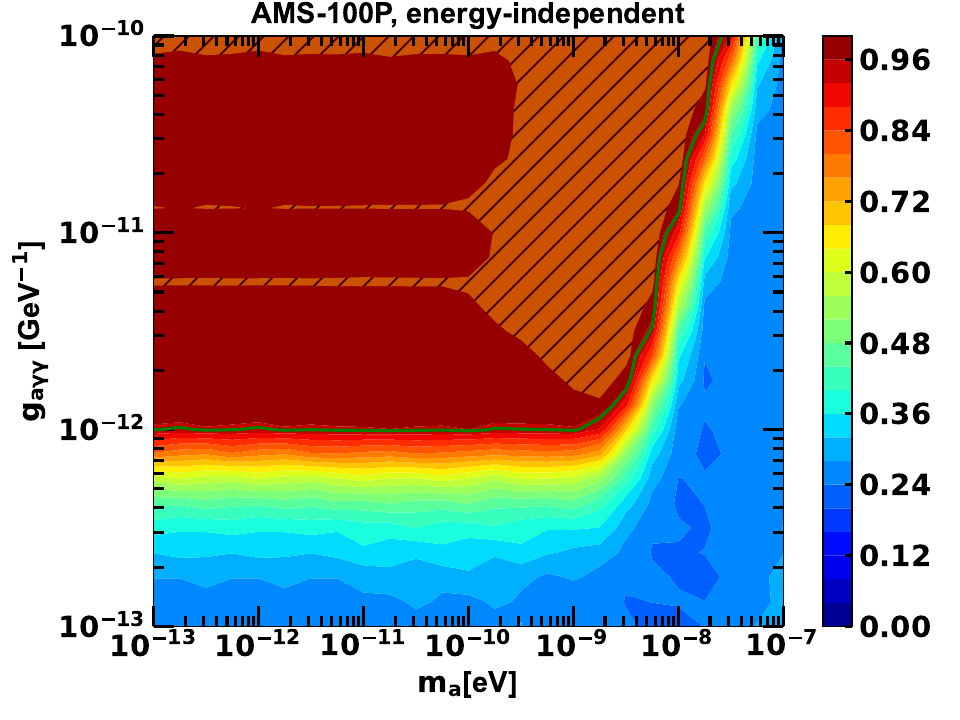} 
  \includegraphics[width=0.49\linewidth]{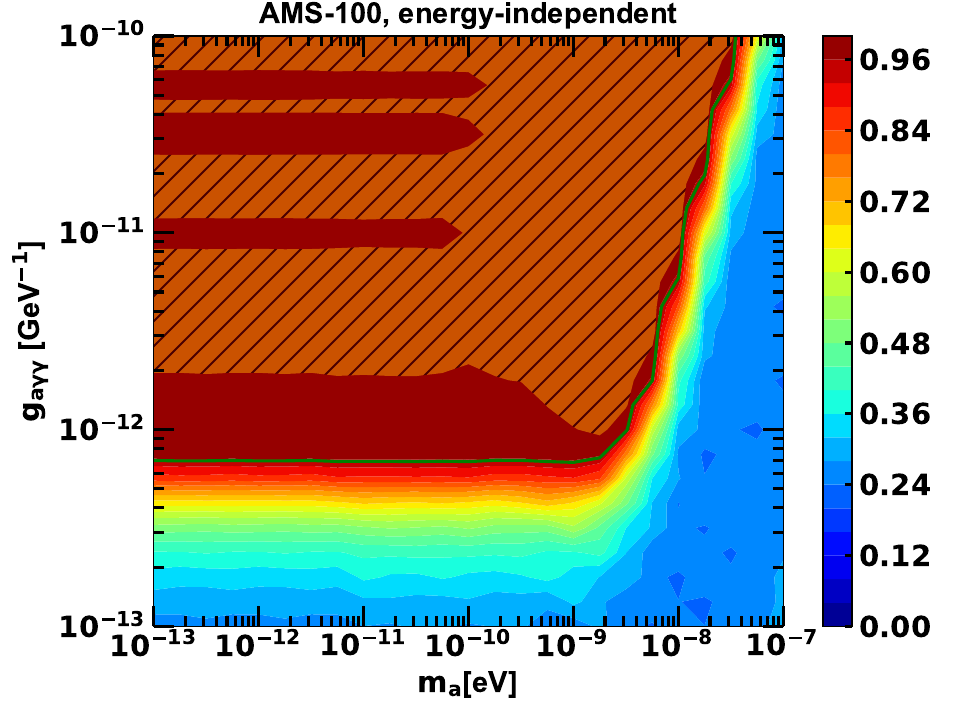}
    \includegraphics[width=0.49\linewidth]{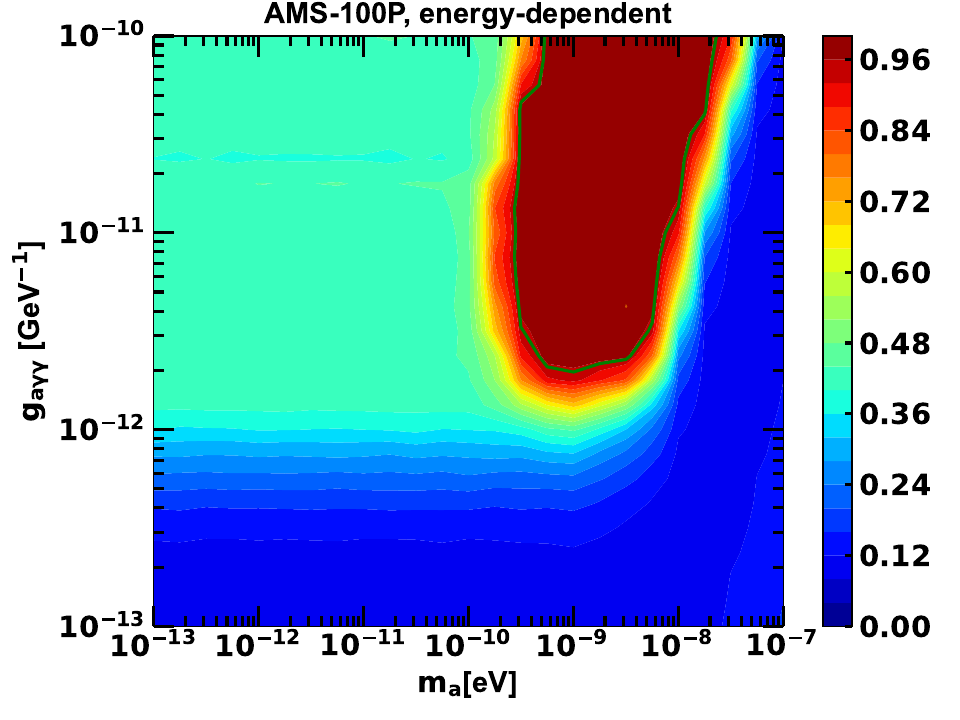} 
  \includegraphics[width=0.49\linewidth]{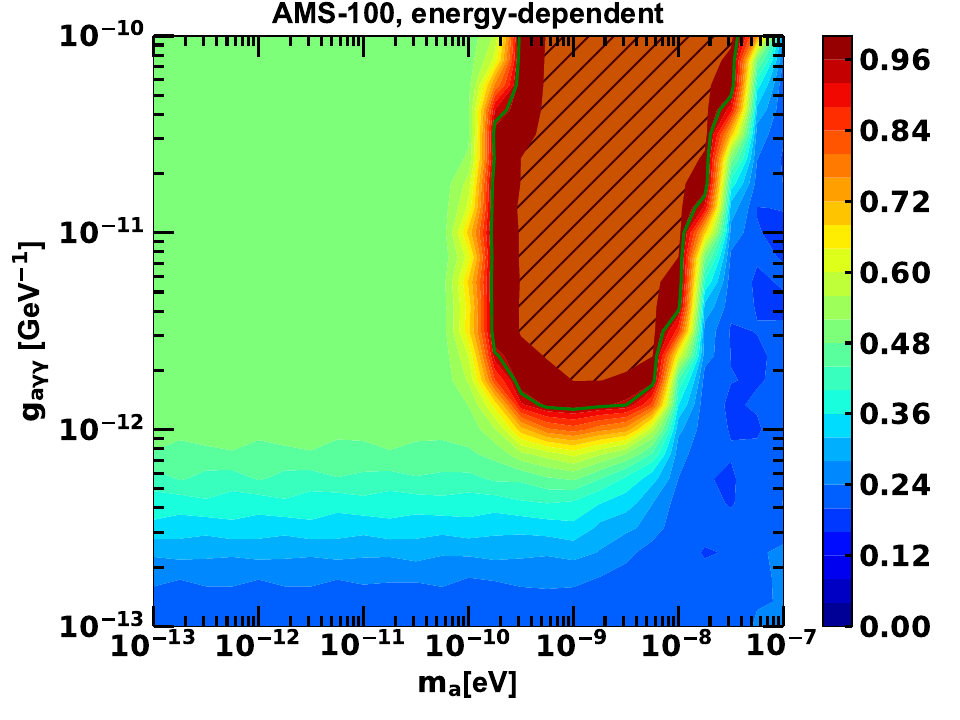}
 \caption{Expected sensitivity with ten years data from NGC1275, for ({\it left panels}) AMS-100P and ({\it right panels}) AMS-100. We randomly sampled 100 $B$-field configurations and calculated the expected $p$-value for each of them; at each point in ALP parameter space we rank the B-field configurations by their corresponding expected $p$-value. All panels show the contours of (1 - expected $p$-value) for ({\it upper panels}) detecting ALP-induced polarization (i.e.~excluding the no-polarization hypothesis), or ({\it lower panels}) detecting energy dependence (i.e.~excluding the constant-polarization hypothesis), for the median B-field scenario at each point. The green line corresponds to the $p=0.05$ contour (for the median B-field scenario). The cross-hatched region in each plot shows the contour corresponding to a $p$-value of 0.05 for the 95th best B-field scenario (out of 100) at each point.
 %In the left panel we also show (dotted line) the contour of expected confidence level $0.95$ for the 5th best B-field scenario (of 100) for each point, corresponding to a favorable $B$-field model. 
%The {\it bottom} panel shows the contours of expected confidence level for detecting polarization in the 95th best B-field scenario (of 100) at each point, for ten years of AMS-100 data, corresponding to an unfavorable $B$-field model. The {\it bottom right panel} shows the contours of expected confidence level for exclusion of the constant-polarization hypothesis, with ten years of AMS-100 data, for the median B-field scenario. 
}
  \label{fig:senEG}
 \end{center}
 \end{figure*}

 We summarize the regions of ALP parameter space where we expect to be able to exclude the relevant null hypotheses, using NGC1275 observations, in Fig.~\ref{fig:senEG}. We show contours of the expected (1 - $p$-value) for exclusion of an unpolarized signal in AMS-100P (upper left) and AMS-100 (upper right), and for exclusion of an energy-independent polarization in the same two experimental configurations (lower left and right). In all panels, we show the results for the median realization (ranked by $p$-value at each point) of the NGC1275 $B$-field. In the cross-hatched regions we show the expected $p=0.05$ contour when instead at each ALP parameter point we choose the $B$-field configuration that is ranked 95th (out of 100) in terms of the expected $p$-value. %Following this procedure for AMS-02 would mean the expected $p$-value is always above 0.05, i.e.~there is no point in ALP parameter space where we expect AMS-02 data to exclude the no-polarization hypothesis at $95\%$ confidence or better  in at least $95\%$ of possible $B$-field configurations.  In the lower right panel, for the energy-dependence analysis with AMS-100, we show the region where the expected confidence level for detection exceeds 0.95 for the ``unfavorable'' (95/100) $B$-field configuration as a cross-hatched overlay on the contours for the median $B$-field realization. 
 We observe that in the analysis of energy-dependence, in particular, the effect of the more unfavorable $B$-field model is quite modest for AMS-100. In the parameter region where we potentially have sensitivity to energy-dependence with AMS-100, high confidence levels for detection are ubiquitous even for quite unfavorable $B$-field models. However, this statement is sensitive to the exact exposure: AMS-100P has almost the same sensitivity as AMS-100 in the presence of the median $B$-field model, but the sensitivity to energy-dependence is almost entirely lost for an unfavorable $B$-field model.

\begin{figure}[htbp]
\begin{center}
  \includegraphics[width=1.05\linewidth]{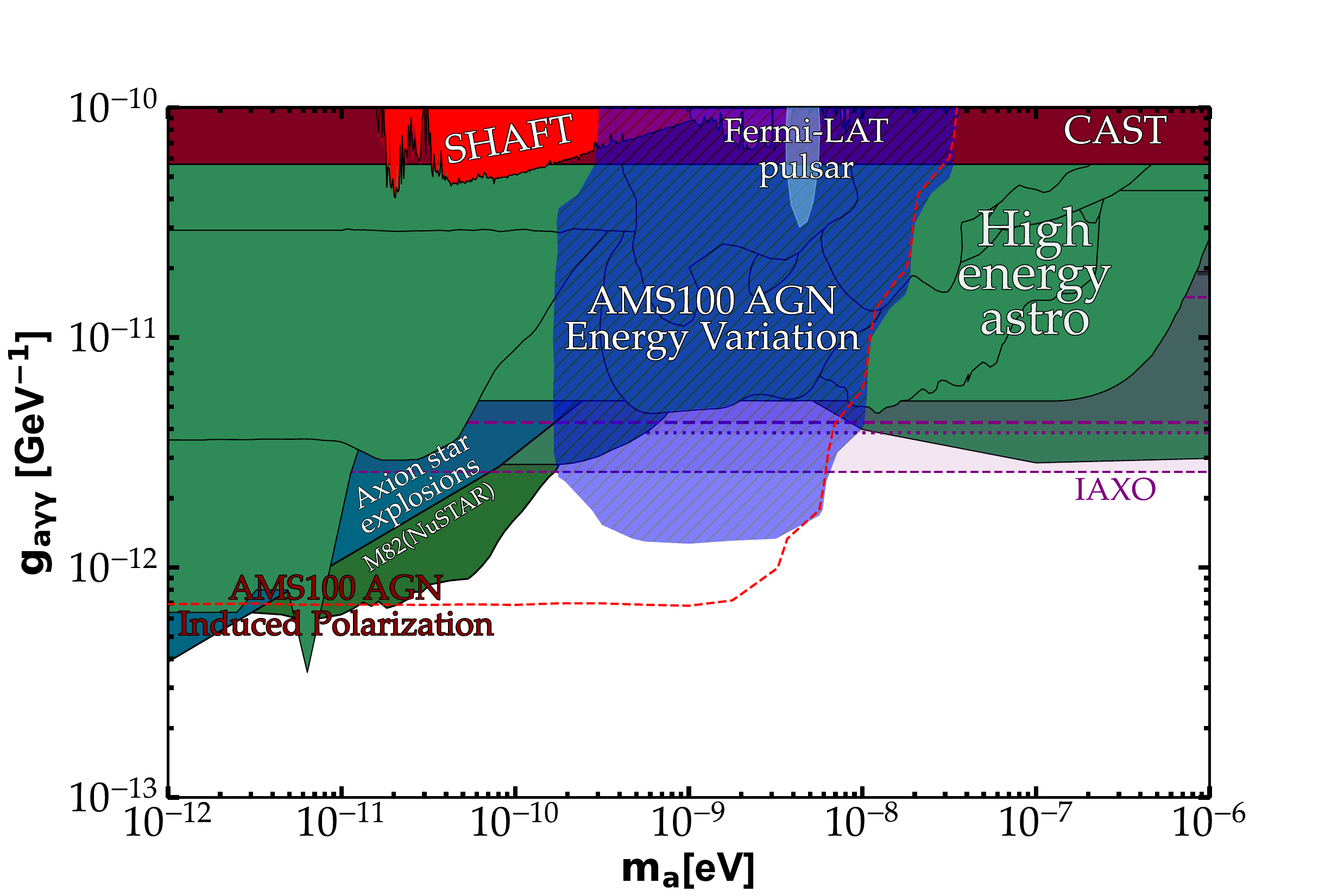}
 \caption{Sensitivity region (expected $p\le 0.05$) for signals from NGC1275 with the median $B$-field configuration, compared to other existing constraints. The shaded region indicates the projected sensitivity region for detecting an energy-dependent polarization signal from AMS-100 with ten years of data. The red dashed lines indicate the projected sensitivity region for detecting an ALP-induced polarization signal with ten years of data from AMS-100, as labeled in the plot. The possible signal from Fermi-LAT~\cite{Majumdar:2018sbv} is shaded in light blue. The other existing constraints are plotted using AxionLimitPlotter~\cite{AxionLimits}.}
  \label{fig:senscompEG}
 \end{center}
 \end{figure}

In Fig.~\ref{fig:senscompEG}, we compare our derived sensitivity regions to existing constraints. 
%We find that, with the existing AMS-02 dataset and for the median $B$-field realization, the possible ALP parameter space that can be probed is already in tension with other existing constraints. For a favorable $B$-field realization, however, it might be possible to observe ALP-sourced polarization from NGC1275 for a region of parameter space that is not yet excluded in the mass range around 0.1-1 neV; for example, we see in the upper left panel of Fig.~\ref{fig:senEG} that for the 5th-best $B$-field model out of the 100 we tested (ranked by expected $p$-value), AMS-02 could have sensitivity to ALP-induced polarization for $g_{a\gamma \gamma} \gtrsim 10^{-12} \text{GeV}^{-1}$ for $m_a \lesssim 2\times 10^{-10}$ eV, with some sensitivity up to $m_a \lesssim 3 \times 10^{-9}$ eV. Of course, such a detection could not be taken as unambiguous evidence of ALPs unless we could place a stringent prior constraint on the degree to which the AGN emission is itself polarized. Furthermore, these more favorable B-field models might impact constraints on ALP-induced signals from the gamma-ray intensity spectrum of NGC1275 with Fermi-LAT data (as in e.g.~\cite{Fermi-LAT:2016nkz, Pallathadka:2020vwu, Cheng:2020bhr}). An analysis using a common set of B-field configurations (for both Fermi-LAT and AMS-02 signals) would be needed to establish whether AMS-02 could hope to see a polarization signal without the ALP explanation being simultaneously ruled out by Fermi-LAT intensity data.
We see that AMS-100, or even a smaller pathfinder such as AMS-100P, would extend its reach into parameter space that is not yet tested by any experiment, and also offer the possibility of detecting energy dependence of the polarization for $10^{-10}$ eV $\lesssim m_a \lesssim 10^{-8}$ eV. The region in which energy dependence could potentially be detected is similar in shape and reach to (but does not perfectly coincide with) the region excluded by updated analyses of gamma-ray intensity variation from NGC1275 using Fermi-LAT data \cite{Pallathadka:2020vwu, Cheng:2020bhr}, with our forecasts for both AMS-100P and AMS-100 extending to slightly lower coupling; this is not surprising, as intensity variations with energy will generally be accompanied by polarization variations, and vice versa. We do not overplot these constraints directly in Fig.~\ref{fig:senscompEG} because the analyses employ different magnetic field models, and this may drive artificial differences between the apparent constraints/sensitivity. A detailed comparison would require use of a shared set of magnetic field configurations.

At lower masses and couplings there is a region of parameter space where there is no detectable energy dependence, but an ALP-induced polarization signal could be measured (by AMS-100, its pathfinder, or a similar experiment). Here there is no competing constraint from gamma-ray intensity measurements, since for intensity measurements the signal of interest intrinsically involves energy-dependent variation.  

%For the polarization-detection analysis, we also tested the sensitivity of a successor experiment with only 10$\times$ the acceptance (such as ALADInO \cite{Battiston2021}), which might be more feasible than the full AMS-100 proposal. The sensitivity to $g_{a\gamma \gamma}$ at small $m_a$ is approximately the geometric mean of the sensitivities for AMS-02 and AMS-100, but we find that as $m_a$ is increased, the cutoff in sensitivity occurs at a similar $m_a$ value to our AMS-100 forecast. Such an experiment would also probe currently unconstrained ALP parameter space for the median $B$-field configuration.

We have focused in this work on detection sensitivity (for polarization and/or energy-dependent polarization), but we can also consider the upper limits that could be placed on ALP parameter space in the event of a null result. Formally, computing the forecast upper limits would require re-simulating our mock data with a different choice of fiducial parameters (i.e.~for the no-signal case, rather than for each choice of ALP parameters); in the energy-dependence analysis this would also entail choosing a fiducial value for the energy-independent polarization fraction. However, to the degree to which the error bars in the inferred polarization (overall or bin-by-bin) are driven by the number of observed photons, the uncertainties will be largely independent of the fiducial choice of polarization model. In this case, we expect a close correspondence between the expected $\chi^2$ test statistic for the signal hypothesis when the ground truth contains no signal, and the test statistic for the case we computed, when the ground truth and the hypothesis being tested are reversed (since this just corresponds to a sign flip in $P_i - P_\text{model}$ in e.g.~Eq.~\ref{eq:chisq2}, and this term is squared). So in this sense we may expect the region of ALP parameter space that can be ruled out by a null result to be similar to the region where we have sensitivity to detect a signal.

However, this argument neglects the systematic uncertainties associated with the modeling of the magnetic field. 
%In particular, for AMS-02, a null result in a polarization search targeting NGC1275 will not imply {\it any} robust exclusion of ALP parameter space. This follows from our result that for AMS-02, for every point in ALP parameter space, in at least $5\%$ of B-field scenarios the expected confidence level for detection is below 95\%. Thus we do not generally expect to be able to argue that a lack of detected polarization in AMS-02 implies the absence of ALPs (at $95\%$ confidence or better), as the absence of a signal could simply be due to an unfavorable $B$-field configuration. 
As discussed above, in the region where AMS-100 would have sensitivity to energy-dependent polarization, the signal is  relatively robust to uncertainties in the $B$-field modeling, and accordingly a null result (either in the overall polarization search or the search for energy-dependence) could provide constraints on ALPs.  The lower-mass region of ALP parameter space where we would expect to detect polarization but no energy-dependence will be more challenging to either distinguish from astrophysics (in the presence of a polarization detection) or to firmly exclude (in the event of a null result), given the dependence on the $B$-field configuration shown in Fig.~\ref{fig:senEG}. The full version of AMS-100 would improve markedly on AMS-100P in this regard, for both energy-independent and energy-dependent analyses.

 \subsection{Galactic sources} 

 \begin{figure*}[htbp]
\begin{center}
  \includegraphics[width=0.49\linewidth]{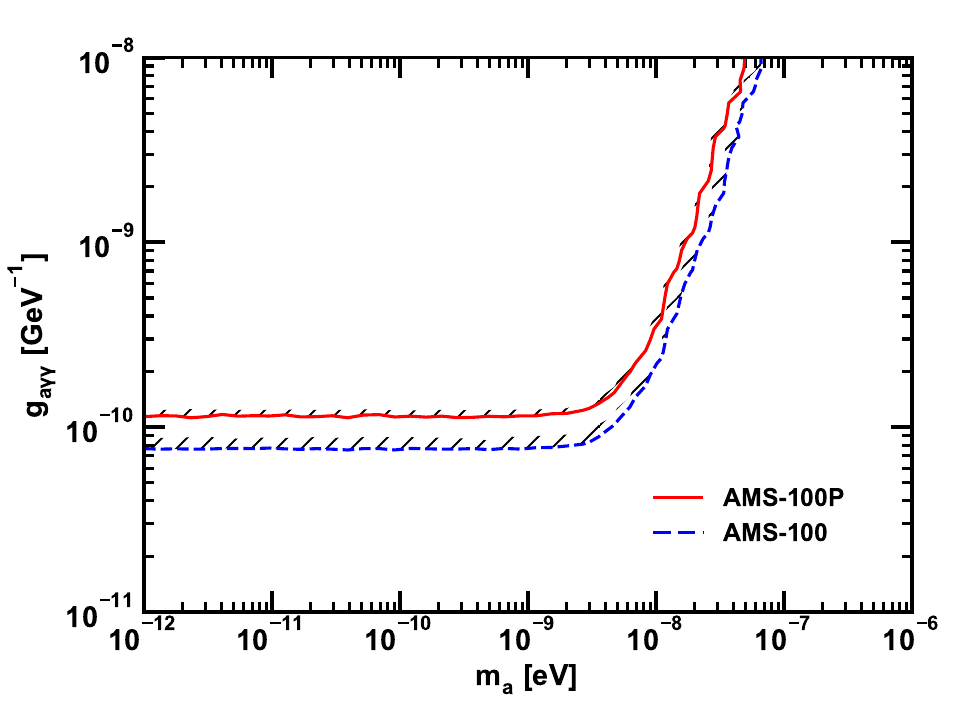}
  \includegraphics[width=0.49\linewidth]{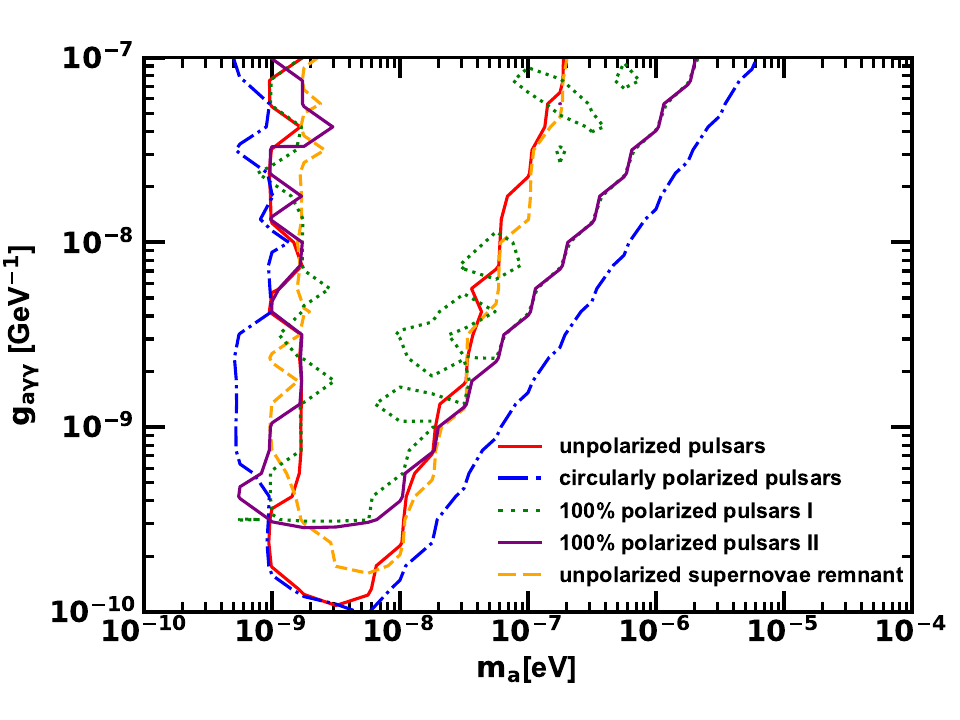} 
 \caption{{\it Left panel:} region of ALP parameter space where the expected $p$-value satisfies $p < 0.05$, for detection of ALP-induced polarization (assuming initially unpolarized sources), using ten years of data from AMS-100P (red) and AMS-100 (blue). The sources employed are four Galactic supernova remnants (see text for details). {\it Right panel:} region of ALP parameter space where the expected $p$-value satisfies $p < 0.05$, for detection of ALP-induced energy-varying polarization, using ten years of data from AMS-100. Different lines correspond to different source populations, plus (in the case of pulsars) different assumptions for the initial polarization. The two cases with $100\%$ initial linear polarization correspond respectively to the polarization at the source being aligned with the transverse Galactic magnetic field (green dotted line) and the angle between the two being $\approx \tan^{-1} 3$ (purple solid line), which maximizes the initial oscillation amplitude.
 }
  \label{fig:sengalactic}
 \end{center}
 \end{figure*}
 
We summarize the regions of ALP parameter space where we expect to be able to exclude the relevant null hypotheses, using observations of Galactic sources, in Fig.~\ref{fig:sengalactic}. 
We show the $95\%$ confidence sensitivity to polarization for AMS-100 and AMS-100P in the left panel, and in the right panel show the sensitivity to energy-dependence of polarization for a range of Galactic sources and assumptions about the initial polarization, for AMS-100. In particular, we find that the contours for pulsars and supernova remnants are rather similar, although unpolarized (or circularly polarized) sources allow for probing slightly lower $g_{a\gamma \gamma}$.

We additionally explored measuring variation in the polarization angle with energy, as well as variation in the polarization fraction. This analysis has some additional subtleties due to the periodicity of the polarization angle, and because a measurement of polarization angle necessarily requires a detection of non-zero polarization fraction. However, this analysis could potentially close sensitivity gaps for cases where the polarization is high but almost energy-independent, which can occur for a source that is initially highly polarized.  In practice, we find that the sensitivity region for this analysis is similar to that in Fig.~\ref{fig:sengalactic}, so we do not show it separately; more specifically, the inclusion of the energy dependence for the polarization angle slightly extends the minimum $g_{a\gamma \gamma}$ that can be probed in  the case of initially linearly polarized sources to be similar to that for unpolarized sources.

  \begin{figure*}[htbp]
\begin{center}
  \includegraphics[width=0.49\linewidth]{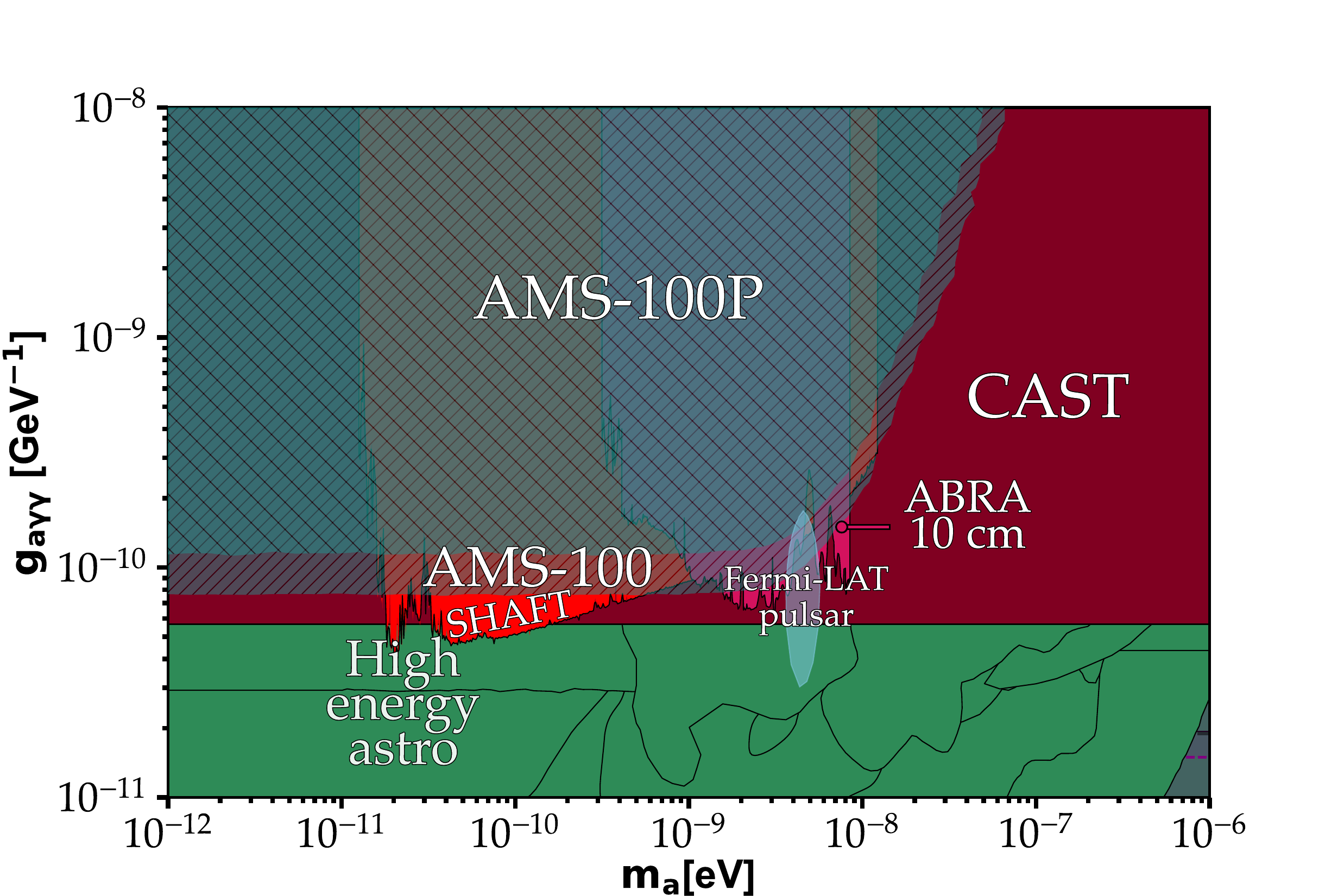} 
  \includegraphics[width=0.49\linewidth]{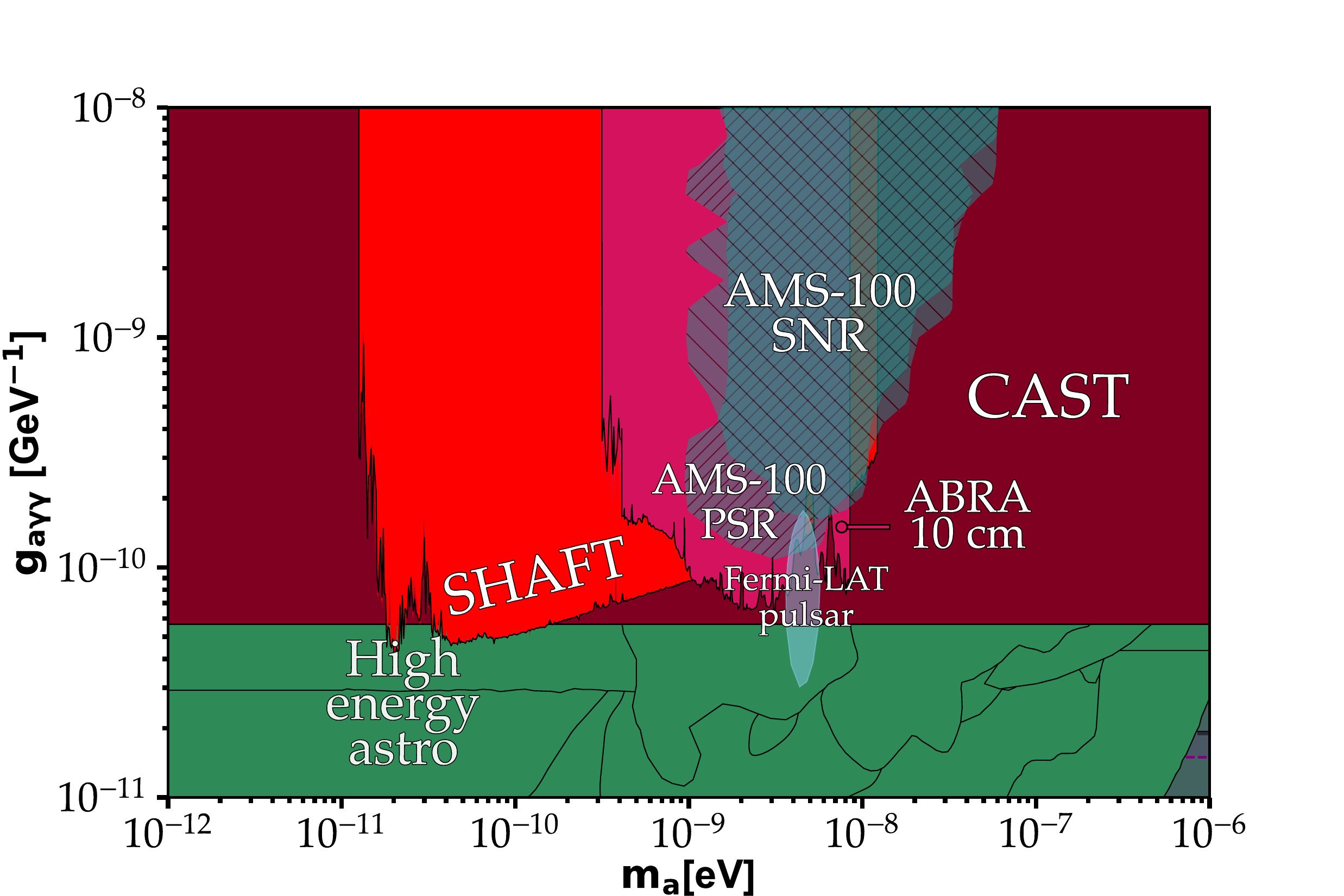}
 \caption{{\it Left panel:} Region with expected $p$-value satisfying $p \le 0.05$ for detection of non-vanishing linear polarization, compared to other existing constraints. {\it Right panel:} Region with expected $p$-value satisfying $p \le 0.05$ for detection of energy dependence in linear polarization, compared to other existing constraints.  Again, the possible signal from Fermi-LAT~\cite{Majumdar:2018sbv} is shaded by light blue. The other existing constraints are plotted using AxionLimitPlotter~\cite{AxionLimits}.}
  \label{fig:senscompG}
 \end{center}
 \end{figure*}

In Fig.~\ref{fig:senscompG}, we compare our derived sensitivity regions for Galactic sources to existing constraints. The results are less optimistic compared to the extragalactic case, with the entire parameter space in which AMS-100 could detect ALP-induced polarization being nominally in conflict with CAST limits (and others). However, there is overlap between the AMS-100 forecast sensitivity and the parameter region proposed in Ref.~\cite{Majumdar:2018sbv} to explain intensity spectrum variations in pulsars observed by Fermi-LAT, suggesting that for specific models that explain this signal while evading the CAST limits (and other competing bounds), a polarization signal could potentially be within reach in the future. 

Similarly, a null result in this search (even with AMS-100) would not nominally constrain new ALP parameter space due to the overlap with existing bounds, but could help test models which evade these bounds due to specific model-dependent features. The uncertainty in the $B$-field modeling is less severe in this case than for the NGC1275 study, but would need to be accounted for in a detailed study of the implications of either a detection or a null result.

  \subsection{Understanding the shape of the constraints}
  
  As expected from the discussion in section \ref{sec:Pheno}, we see that for both Galactic and extragalactic sources, the parameter region where we have sensitivity to ALP-induced polarization requires exceeding a fixed value of $g_{a\gamma \gamma}$ for low $m_A$, then above a threshold value of $m_A$, the limiting value of $g_{a\gamma \gamma}$ rises roughly quadratically with $m_A$. This is consistent with our discussion in section \ref{sec:Pheno}. To exclude energy-independent polarization, we also need to be above a critical mass threshold, which is nearly independent of $g_{a\gamma \gamma}$: this can be understood from our discussion in section \ref{sec:Pheno} suggesting that the low-$m_a$ boundary line for this analysis should scale as $g_{a\gamma \gamma}^{1/4}$.
 
The minimum $g_{a\gamma\gamma}$ to which we have sensitivity is about two orders of magnitude lower for NGC1275 compared to the Galactic source analysis. This can be understood from the fact that in our modeling for NGC1275, the distance traveled in the magnetic field is 1 Mpc, which is around two orders of magnitudes longer than for the Galactic sources. Recall from the discussion in section \ref{sec:Pheno} that for $E \gg E_c$ the conversion probability is controlled by $l_\text{osc}/d$ where $l_\text{osc}\propto 1/g_{a\gamma \gamma}$, so we would naively expect an increase in $d$ by two orders of magnitude to extend sensitivity in $g_{a \gamma \gamma}$ by the same degree (if the $B$-fields are comparable), as observed.

\subsection{\label{supernova} The possibility of a Galactic supernova}

We have focused in this article on steady-state sources. 
It has recently been pointed out that ALP production inside a Galactic supernova could lead to a burst of ALPs escaping prior to the bulk of the electromagnetic radiation (similar to neutrinos), and the conversion of these axions to photons in the magnetic field surrounding the star and in the GMF could lead to a striking signal in gamma-ray telescopes \cite{Manzari:2024jns}. We would expect these converted photons to be  close to fully polarized. In this subsection we briefly estimate how such a signal would appear in AMS-02 polarization measurements.

The field of view for AMS-02 conversion photons can be estimated as 0.5-0.7 sr (depending mildly on energy), from the results for acceptance and effective area provided in Ref.~\cite{Beischer:2020rts}, so the chance that a given Galactic supernova occurs within the field of view is only $\mathcal{O}(5\%)$ (we caution that this is only an order-of-magnitude estimate that does not account for AMS-02's non-uniform exposure). Nonetheless, we will here take the optimistic approach of assuming a Galactic supernova at a distance of 10 kpc, located such that its photons reach AMS-02 with near-perpendicular incidence. To estimate the counts from such an event, we employ the example axion spectrum arising from pions (which is the dominant contribution), integrated for 10s after the supernova, calculated by Ref.~\cite{Manzari:2024jns} for $g_{a\gamma \gamma} = 10^{-12} \text{GeV}^{-1}$ and $E=0.1-1$ GeV. We weight this spectrum by the effective area for AMS-02 pair-conversion photons given in Ref.~\cite{Beischer:2020rts}, for $E=0.2-1$ GeV (the effective area falls rapidly to zero below 0.2 GeV). We find that with the ``KSVZ'' parameters of Ref.~\cite{Manzari:2024jns}, we would expect roughly $9.5\times 10^{10} (g_{a\gamma \gamma}/10^{-12} \text{GeV})^2$ {\it axions} incident on the detector; the measurable photon counts will be this number multiplied by the axion-photon conversion probability. For the ``ALP'' parameters of that reference (differing from the``KSVZ'' case in the couplings relevant for axion production in the supernova), the corresponding number is roughly $26000 (g_{a\gamma \gamma}/10^{-12} \text{GeV})^2$ axions.

Ref.~\cite{Manzari:2024jns} estimate that the conversion probability is $\sim 10^{-5} (g_{a\gamma \gamma}/10^{-12} \text{GeV})^2$ in the GMF, for $m_a \lesssim 2 \times 10^{-11}$ eV. For conversion in the fields of the star, we employ the result given in the supplemental material of Ref.~\cite{Manzari:2024jns} for the red supergiant conversion probability; for energies exceeding 100 MeV, the conversion probability is found in that work to be nearly energy-independent and $\sim 2\times 10^{-8} (g_{a\gamma \gamma}/10^{-12} \text{GeV})^2$, for $m_a \lesssim 5\times 10^{-5}$ eV.

For a blue supergiant progenitor, the conversion probability is estimated by the authors of Ref.~\cite{Manzari:2024jns} to be $\sim 6\times 10^{-9}$ for $g_{a\gamma\gamma}=10^{-12} \text{GeV}^{-1}$ and a surface magnetic field of 100G, corresponding to the conservative constraints in that work. For a surface magnetic field of 1 kG, which the authors of that work consider more realistic, the conversion probability (for $g_{a\gamma\gamma}=10^{-12} \text{GeV}^{-1}$) varies from $\sim 6\times 10^{-7}$ at 0.1 GeV to $\sim 3 \times 10^{-7}$ at 1 GeV, corresponding to a signal roughly one order of magnitude larger than the red supergiant case \cite{privcommsafdi}.

With the red supergiant conversion probabilities, we see that in order to distinguish a fully polarized signal from an unpolarized signal at 95\% confidence (i.e. requiring MDP $< 1$, so at least $\sim 600$ counts), for AMS-02 we would need $g_{a\gamma \gamma} \gtrsim 7 \times 10^{-12}$ GeV$^{-1}$, for the ALP benchmark and for masses $m_a \lesssim 2\times 10^{-11}$ eV (where the conversion in galactic fields dominates). For larger masses, up to $m_a \lesssim 5\times 10^{-5}$ eV, the corresponding sensitivity would be $g_{a\gamma \gamma} \gtrsim 3 \times 10^{-11}$ GeV$^{-1}$. For the KSVZ benchmark, the corresponding numbers are $g_{a\gamma \gamma} \gtrsim 2 \times 10^{-13}$ GeV$^{-1}$ ($m_a \lesssim 2\times 10^{-11}$ eV) and $g_{a\gamma \gamma} \gtrsim 7\times 10^{-13}$ GeV$^{-1}$ ($m_a \lesssim 5\times 10^{-5}$ eV).

AMS-100 is expected to have a field of view close to $4\pi$, thus greatly improving the odds of seeing the supernova in the first place. If we approximate its effective area as the effective acceptance divided by $4\pi$, we would obtain a peak effective area of $30/4\pi \approx 2.4$ m$^2$, compared to $180$ cm$^2$ for AMS-02. This corresponds to an increase in effective area by a factor of around 130.

Since the axion luminosity and the conversion probability each scale independently as $g_{a\gamma\gamma}^2$, an AMS-100-like experiment would improve the sensitivity to $g_{a\gamma\gamma}$ by a factor of $130^{1/4}\approx 3.4$. Similarly, if the stellar conversion probability we use above is underestimated by roughly an order of magnitude (as might be the case for a blue supergiant progenitor with kG surface magnetic field), that would improve the sensitivity to $g_{a\gamma\gamma}$ by a factor of $10^{1/4}\approx 1.8$ in the high-ALP-mass region where conversion on stellar magnetic fields dominates.

\section{\label{results}Summary and Outlook}

We have analyzed the detectability of gamma-ray polarization, both astrophysical and induced by new physics such as ALP-photon mixing, in the AMS-02 and (proposed) AMS-100 detectors. Compared with existing X-ray polarization measurements, gamma-ray polarization would both provide a new window on astrophysical production mechanisms for gamma-rays in bright sources, and could offer sensitivity to regions of ALP parameter space where the critical energy for oscillations is high and consequently the signal is suppressed at low photon energies.

We have shown that with AMS-02, we only expect sufficient statistics to detect polarization for the brightest Galactic gamma-ray sources, such as the Vela and Geminga pulsars. These sources are relatively close to Earth and we do not expect a significant ALP-induced polarization signal, but an indication of any gamma-ray polarization in these sources could shed light on their (astrophysical) gamma-ray production mechanisms and serve as a proof of principle for future searches. 

Detection of ALP-induced polarization in gamma rays would thus have to wait for a future mission, but we have shown that AMS-100, or a proposed pathfinder with smaller effective area by a factor of 5 (denoted AMS-100P), would each be able to probe currently unconstrained regions of ALP parameter space using observations of the bright AGN NGC1275. Specifically, AMS-100 would extend the $g_{a\gamma\gamma}$ reach down to $\sim 7\times 10^{-13}$ GeV$^{-1}$ for masses up to $\sim 2 \times 10^{-9}$ eV, with the median $B$-field configuration, and would retain much of this sensitivity even for unfavorable $B$-field configurations (at the $5$th percentile of significance, among $B$-field models we tested).

Galactic sources are less favorable for probing small $g_{a\gamma \gamma}$ due to the smaller propagation distance through substantial $B$-fields (compared to NGC1275, which is located in the core of the Perseus cluster). We have shown that the region where AMS-100 has sensitivity is nominally excluded by existing bounds from CAST, SHAFT, and ABRA 10cm. This is broadly consistent with previous studies of energy-dependent modulation of the intensity signal in gamma-rays from pulsars, which claimed a nominal detection of a signal, but in ALP parameter space that is in tension with existing bounds \cite{Majumdar:2018sbv}. Some models have been proposed to alleviate or remove these tensions, and in such models, the AMS-100 sensitivity region from our work overlaps with the signal region found in the literature for intensity fluctuations, and could thus serve as an independent check for at least part of the parameter space. Similarly, gamma-ray observations of a Galactic supernova could offer sensitivity to new regions of ALP parameter space \cite{Manzari:2024jns}, and we have estimated the sensitivity of AMS-02 to the polarization signal from such a source (assuming the roughly $5\%$ chance that the supernova occurs within the AMS-02 field of view). AMS-100 would be an excellent supernova detector due to its nearly $4\pi$ field of view, and we have also estimated the sensitivity in this case. Polarization is unlikely to be the first detection channel for ALPs in such an event, as a polarization measurement requires a substantial photon flux (in AMS-like detectors, at least 600 photons are required for a fully polarized signal to be above the MDP with $p=0.05$), but could serve as a powerful cross-check on the ALP origin of such a signal.

\section*{Acknowledgements}

We thank the anonymous referee for helping us uncover an error in the AMS-02 acceptance modeling in an earlier version of this work. We thank Francesca Calore and Ben Safdi for helpful conversations, and Carlos Ma\~{n}a Barrera for his work on measuring polarization in AMS-02. YJ is grateful for the support of  DOE including resources from the National Energy Research Scientific Computing Center under Contract No. DE-AC02- 05CH11231. XZ is supported by DOE grant number DE-SC0024112. TRS was supported in part by a Guggenheim Fellowship; the Edward, Frances, and Shirley B. Daniels Fellowship of the Harvard Radcliffe Institute; the Bershadsky Distinguished Fellowship of the Harvard Physics Department; and the Simons Foundation (Grant Number 929255, T.R.S). TRS thanks the Kavli Institute for Theoretical Physics (KITP), the Aspen Center for Physics, and the Mainz Institute for Theoretical Physics for their hospitality during the completion of this work; this research was supported in part by grant no.~NSF PHY-2309135 to KITP, and performed in part at the Aspen Center for Physics, which is supported by National Science Foundation grant PHY-2210452. This work was supported by the U.S. Department of Energy, Office of Science, Office of High Energy Physics of U.S. Department of Energy under grant Contract Number DE-SC0012567. 

\appendix

\section{Note on choice of energy binning}
\label{app:binning}

One of the choices we made in this study is the number of energy bins characterizing the energy dependence. In general, increasing the number of bins is preferable to better capture the oscillation pattern, but is computationally more expensive; very narrow bins may also violate our large-photon-count approximations and require us to use the full Poisson likelihood. By default, we use four energy bins in our Galactic analysis and ten energy bins in our extragalactic (NGC1275) analysis. Here in this section we discuss the possible effects of choice of binnings on our result and justify our choices.

We do not expect this choice to significantly affect the minimum $g_{a\gamma \gamma}$ to which we have sensitivity; as described in section \ref{sec:Pheno}, this boundary is primarily determined by the oscillation length $l_\text{osc}$. However, using a too-small number of bins could lead to ``holes'' in our sensitivity region, due to averaging out of oscillations over too-wide bins, and could also lead to a loss of sensitivity at lower $m_a$ where the critical energy $E_c$, and hence the characteristic scale of the oscillations, is reduced. This is especially the case for the extragalactic analysis, where the uncertainty in the $B$-field means that it is possible to find $B$-field configurations that evade detection specifically by falling into one of these averaging-induced blind spots. 

\begin{figure*}[htbp]
\begin{center}
  \includegraphics[width=0.49\linewidth]{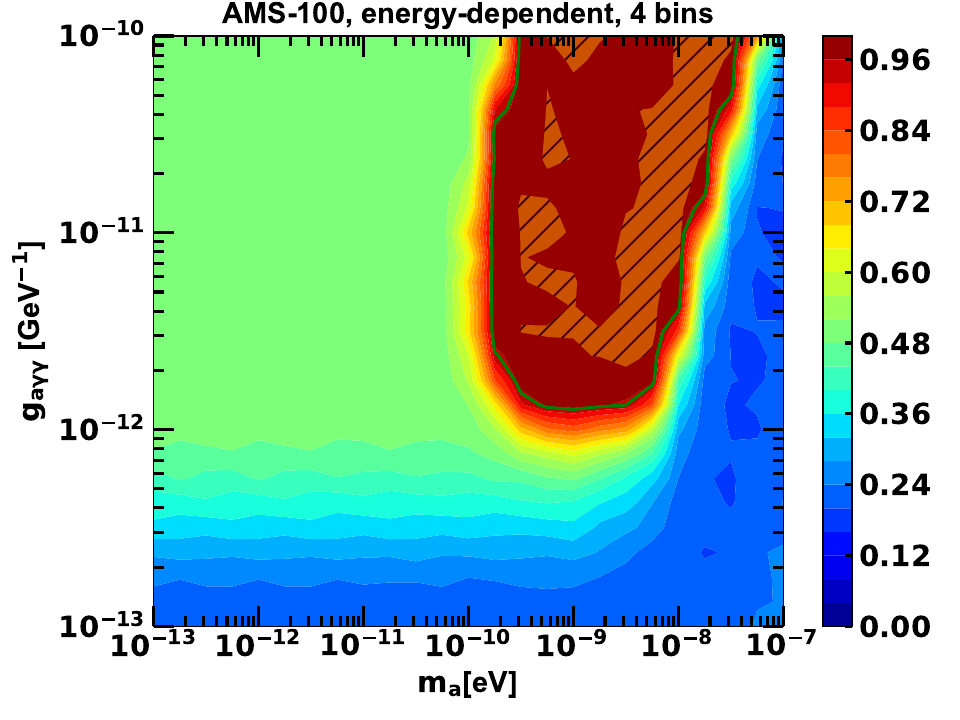} 
  \includegraphics[width=0.49\linewidth]{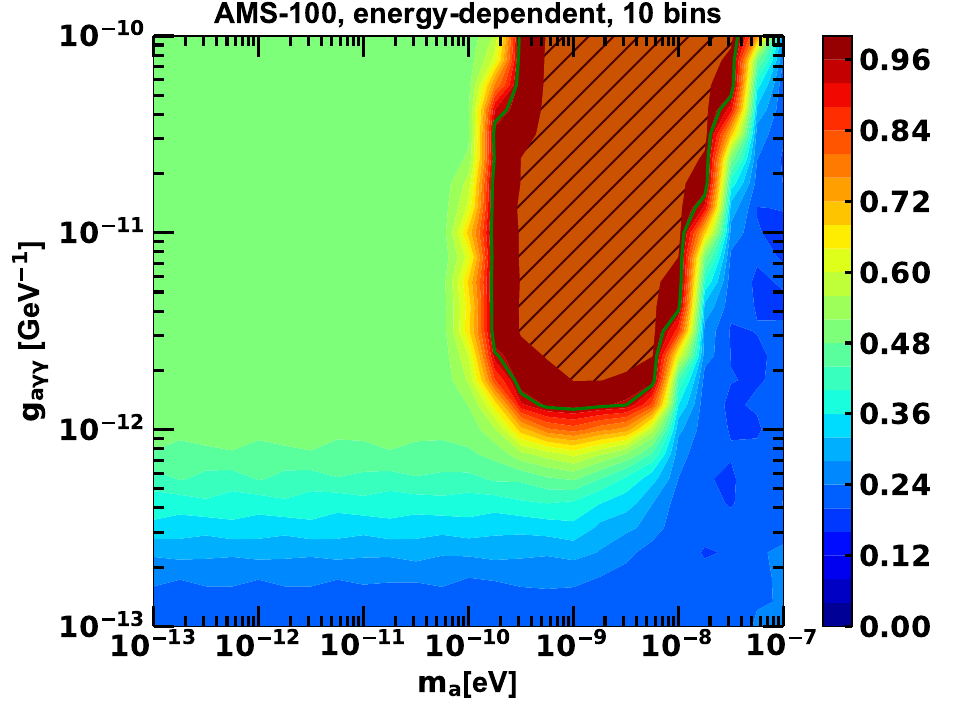}
 \caption{Sensitivity (1 - expected $p$-value) to energy-dependence of polarization with ten years  of AMS-100 data from NGC1275. Contours show the results for the median $B$-field configuration (ranked by expected $p$-value); the cross-hatched region corresponds to  $p \le 0.05$ for an unfavorable $B$-field model (95th/100 in significance).  {\it Left panel:} results of analysis with 4 log-spaced energy bins. {\it Right panel:} results of analysis with 10 log-spaced energy bins, as in Fig.~\ref{fig:senEG}.}
  \label{fig:binningstudy}
 \end{center}
 \end{figure*}

We demonstrate this point in Fig.~\ref{fig:binningstudy}, which shows the sensitivity regions for the energy-dependence analysis for NGC1275 for the cases of 10 bins and 4 bins. %We see that for galactic sources, adding more energy bins does nothing more than slightly filling the ``holes'' in sensitivity.  
We see that when the median-$p$-value $B$-field configuration is chosen for each parameter point (underlying contours), the binning makes very little difference. But when we instead choose an ``unfavorable'' $B$-field configuration (ranked 95th out of 100 in terms of $p$-value at that point), using only 4 bins removes an $\mathcal{O}(1)$ fraction of the region where the expected $p$-value is below 0.05, in particular erasing sensitivity at lower $m_a$, since the ``holes'' in sensitivity are exactly what dominates the ````unfavorable'' case; thus a finer binning improves sensitivity quite considerably. %This justifies our choice to use only 4 bins for the Galactic sources for the sake of computing resources, but 10 bins for NGC1275 to compensate for the large uncertainties coming from our ignorance of the B-field configuration. 
In a real analysis, where there is only one dataset (rather than a large ensemble of simulations), an even finer binning (limited by the AMS-02 energy resolution) might be appropriate.

 % As shown in Fig.~\ref{fig:binningstudy}, the 95CL contour is almost the same for either case in the median realization. However, in the extremity that the odd is against us as in the 95th worst scenario, 10-bin analysis would help capture much more information on the oscillation pattern and thus allow a larger sensitivity region. So more bins are always preferred given enough computational resources and event counts. So in this work we primarily use 4-bin analysis to show the shape of the sensitivity region unless specified otherwise.

    \begin{figure*}[htbp]
\begin{center}
  \includegraphics[width=0.49\linewidth]{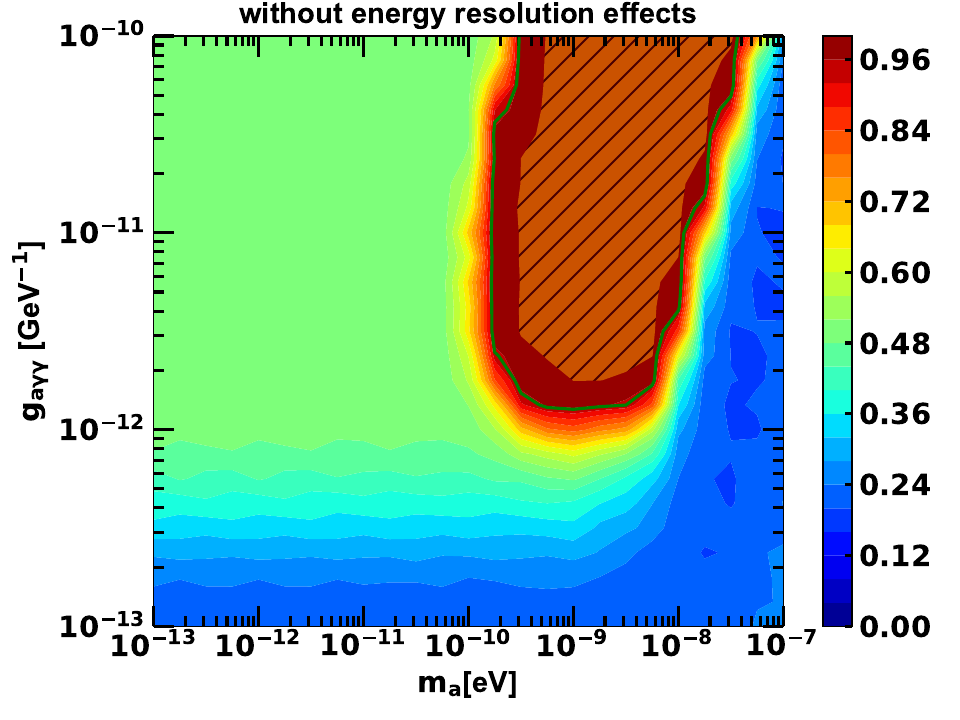} 
  \includegraphics[width=0.49\linewidth]{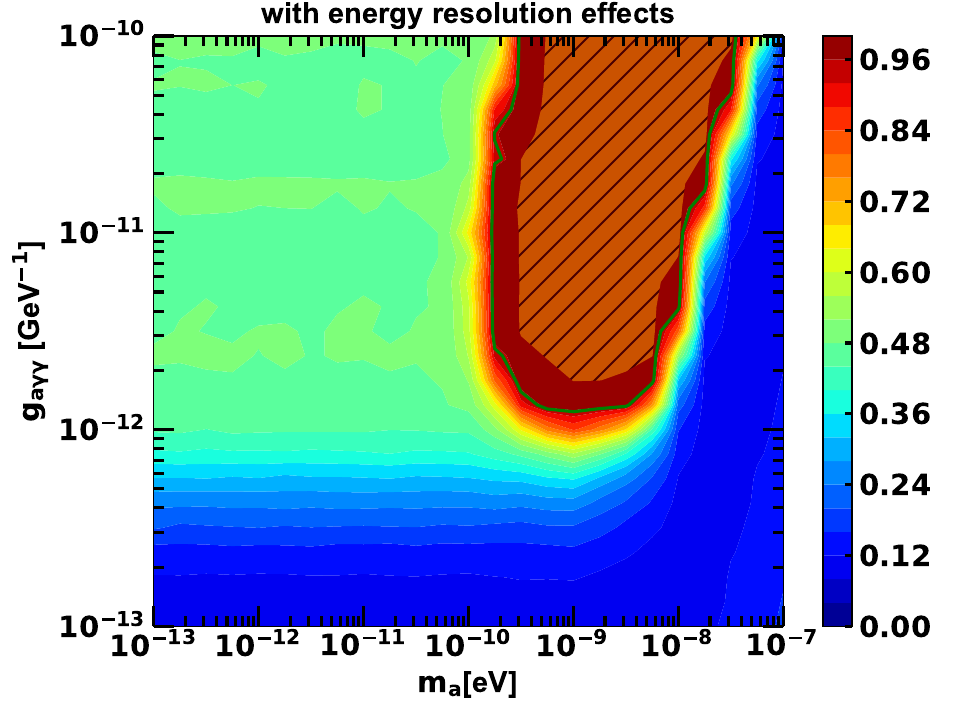}
 \caption{Sensitivity  to energy-dependence of polarization with 10 years of AMS-100 data from NGC1275, as in the bottom right panel of Fig.~\ref{fig:senEG}. {\it Left panel:} contours of $1 - $ expected $p$-value, for the median $B$-field model. The green line corresponds to the expected $p=0.05$ contour. The cross-hatched region corresponds to expected $p \le 0.05$ for the 95th best $B$-field scenario (out of 100). {\it Right panel:} results of identical analysis accounting for energy resolution effects.}
  \label{fig:EnergyResolution}
 \end{center}
 \end{figure*}

However, as shown in ~\cite{Beischer:2020rts}, the energy resolution of AMS-02 is non-negligible: around $15\%$ at 200MeV and grows with energy to $28\%$ at 10 GeV. One might worry that the non-negligible energy resolution will cause a substantial number of photons to be detected in neighboring bins that do not reflect their true energy, thus flattening out the energy dependent polarization pattern, especially for finer binnings. We test this effect explicitly for NGC1275 with 10 bins by assuming an unfavorable $28\%$ resolution for all energy bins. In each energy bin, we randomly draw a thousand samples according to the emission spectrum and, for each sample, we assume the detected photon will follow a Gaussian distribution with a mean value of its true energy and a standard deviation of $28\%$ of the mean value to determine the observed energy. This allows us to model what percentage of photons are transferred to any other bins, as a function of energy. In each energy bin, we average over the polarization as before but now taking into account the transferred photons as well. From there we redo the rest of the analysis as before with the same set of random $B$-field configurations. The result compared to the one neglecting energy-resolution effects is shown in Fig.~\ref{fig:EnergyResolution}. The plot looks almost identical for $p \lesssim 0.5$, justifying our decision to neglect energy-resolution effects for our default binning (the case with 4 wider bins, rather than 10, will be even less sensitive to the energy resolution), although there is some modification to the contours for higher $p$ (but this in any case corresponds to scenarios where the ALP-induced signal cannot be distinguished from constant polarization).

%\section{Impact of including or omitting Galactic sources}
%\label{app:multisource}
\section{Impact of astrophysical backgrounds}
\label{app:background}

As discussed in the main text, astrophysical diffuse gamma-ray backgrounds may confuse the signal from the sources. In this appendix, we test the effects of backgrounds by adding them to our simulated datasets for AMS-100, assuming that the ALP-induced polarization effect on the backgrounds can be neglected. For each source, we compute the number of expected background counts per bin in energy and azimuthal angle, as described in Sec.~\ref{genericpolsearch}, and then perform a Poisson draw to determine the actual number of background events in each bin in the mock data. We then compute the expected $p$-values (testing for a polarization signal, and for energy-dependent polarization) from these realizations as previously.

   \begin{figure*}[htbp]
\begin{center}
  
    \includegraphics[width=0.49\linewidth]{Fig9_sen1.pdf}
\includegraphics[width=0.49\linewidth]{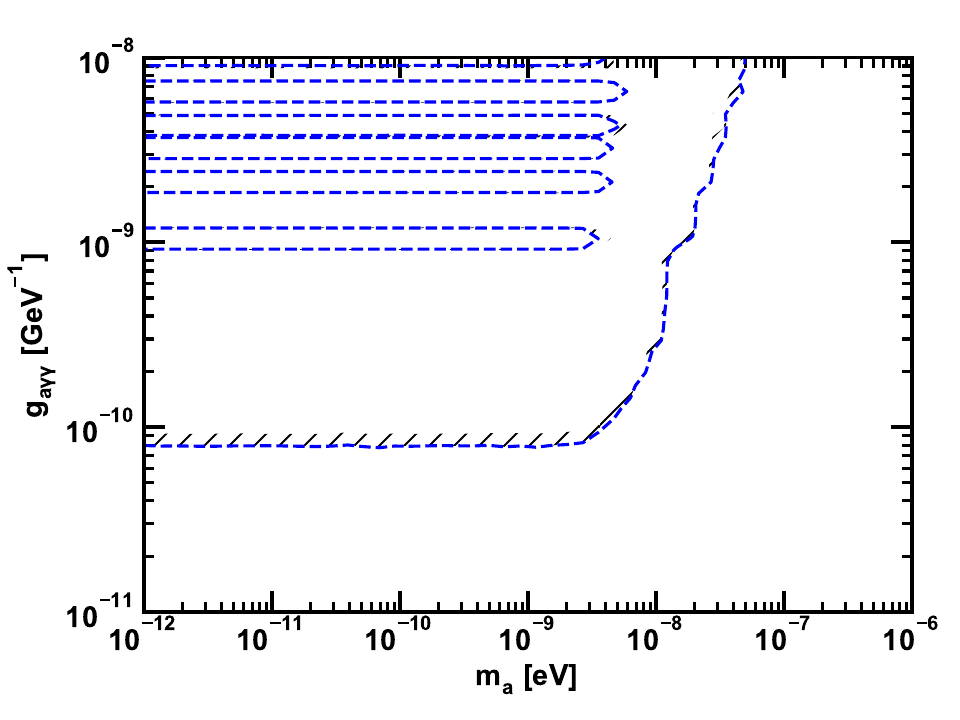}
  \includegraphics[width=0.49\linewidth]{Fig9_sen2.pdf} 
  \includegraphics[width=0.49\linewidth]{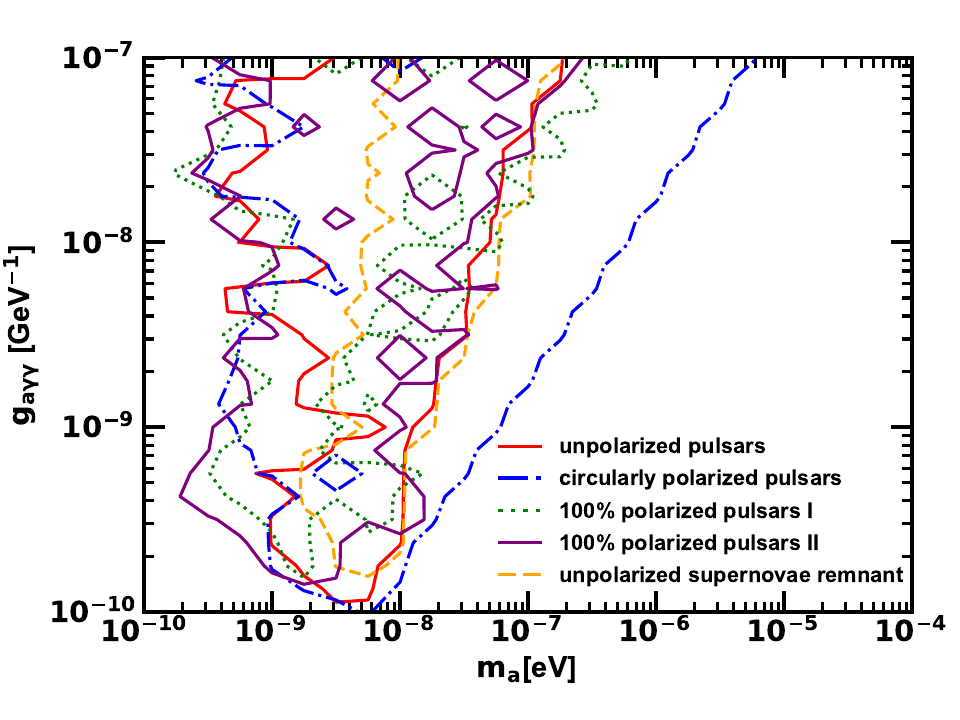}
 \caption{ Contours of $p=0.05$ for testing the presence of polarization ({\it upper panels}) and the energy-dependence of polarization ({\it lower panels}) with 10 years of AMS-100 data from Galactic sources. {\it Left panels} show the results with no background, as in the left and right panels of Fig.~\ref{fig:sengalactic}. {\it Right panels} show the results of the identical analysis with background included (in the upper right panel, results are for AMS-100 only).}
  \label{fig:bkgstudygalactic}
 \end{center}
 \end{figure*}

\begin{figure*}[htbp]
\begin{center}
  \includegraphics[width=0.49\linewidth]{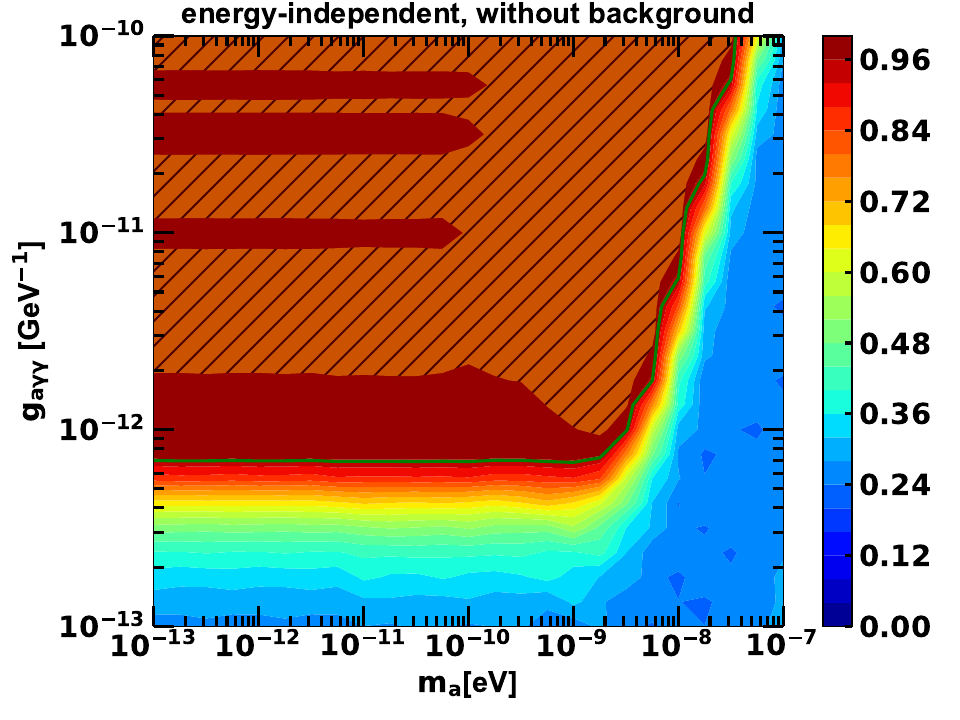}
  \includegraphics[width=0.49\linewidth]{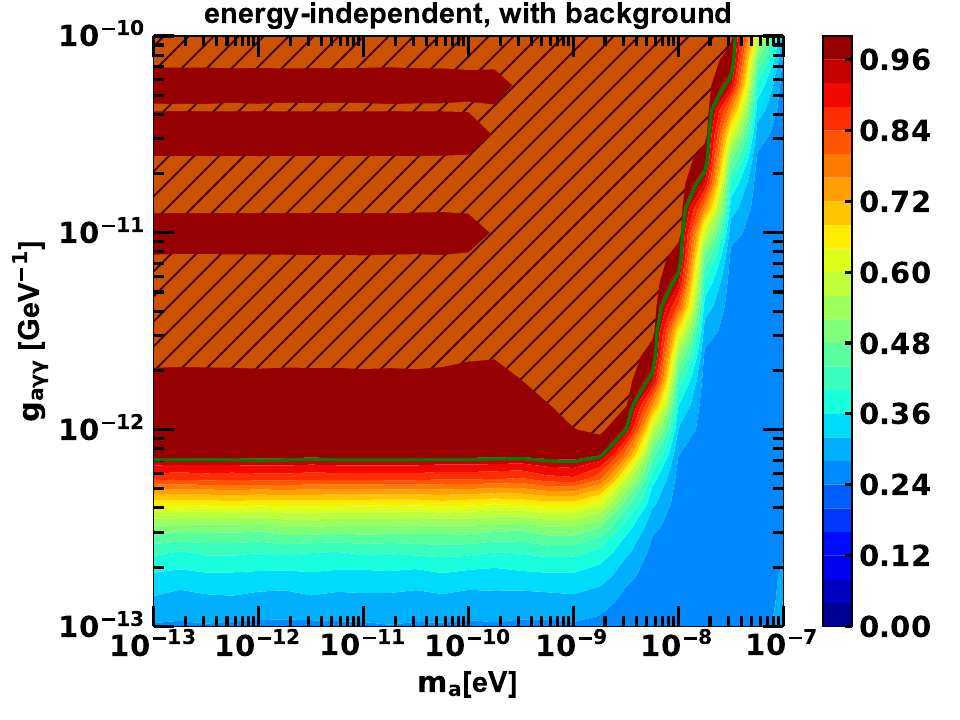}
  \includegraphics[width=0.49\linewidth]{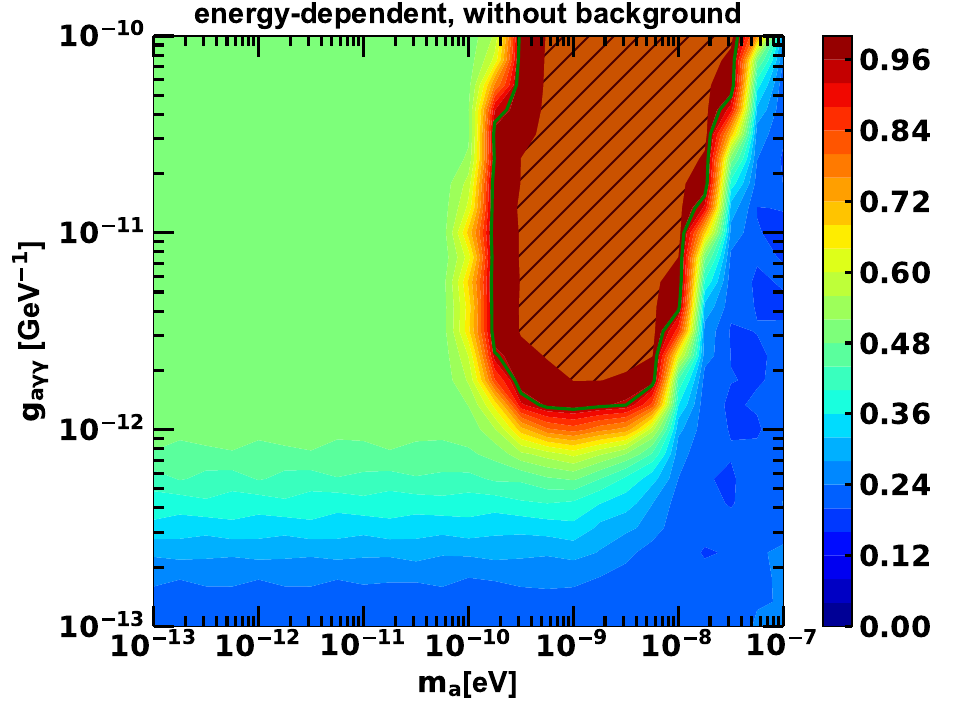} 
  \includegraphics[width=0.49\linewidth]{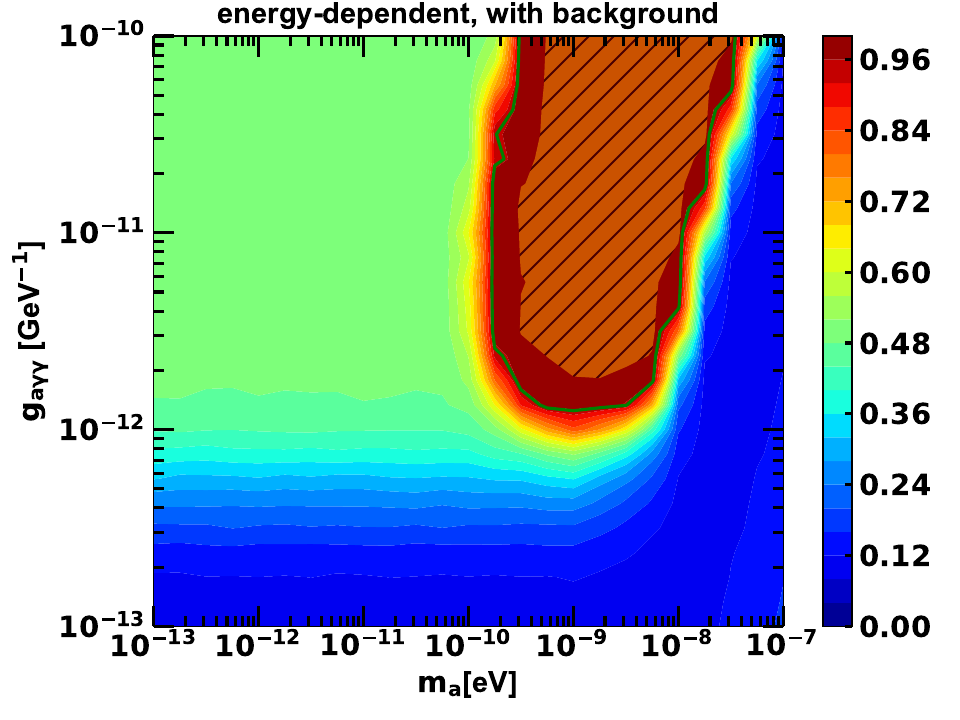}
 \caption{Sensitivity to polarization ({\it upper panels}) or energy-dependence of polarization ({\it lower panels}) with 10 years of AMS-100 data from NGC1275. {\it Left panels:} contours of $1 - $ expected $p$-value, for the median $B$-field model, with no background, as in the right panels of Fig.~\ref{fig:senEG}. The green line corresponds to the expected $p=0.05$ contour. The cross-hatched region corresponds to expected $p \le 0.05$ for the 95th best $B$-field scenario (out of 100). {\it Right panels:} results of identical analysis accounting for backgrounds.}
  \label{fig:bestsource}
 \end{center}
 \end{figure*}

Fig.~\ref{fig:bkgstudygalactic} shows the impact of this choice on the analyses searching for ALP-induced polarization from Galactic sources. We observe that inclusion of backgrounds noticeably degrades the sensitivity region, largely by removing ``islands'' of lower sensitivity rather than by changing the minimum testable $g_{a\gamma \gamma}$. Fig.~\ref{fig:bestsource} performs the analogous study for NGC1275, where we see there is very little impact on the $p \le 0.05$ region (although there is some impact on the shape of the contours for higher $p$ in the energy-dependent analysis). This makes sense as the diffuse background is relatively much larger for distant (and therefore fainter) Galactic sources that lie in the Galactic plane, compared to NGC1275, which is both bright and located away from the Galactic plane.

% For Galactic sources, combining polarization detections from multiple sources could increase the significance of any signal and serve as a non-trivial test for the ALP hypothesis. While for any one source the Earth may accidentally lie at a point where the ALP-induced polarization is small (since the fraction of converted photons is an oscillatory function of distance), it is much less likely that the polarization would be accidentally small for multiple sources at different distances. For this reason, combining sources can help remove islands in parameter space where an ALP signal is undetectable for a specific source.

% In Fig.~\ref{fig:bestsource} we show the expected region for sensitivity to energy-dependence in the polarization when only a single source is included (for the sources with the highest photon flux), and also for the combined results with that one source removed.

% We see that the shape of the constraint region is in both cases generally similar to that of the combined analysis, implying that the signal is not dominated by one source. However, omitting J2021+3651 modestly increases the minimum $g_{a\gamma \gamma}$ that can be probed by the pulsar ensemble, and decreases the mass reach for the (perhaps unphysical) case of a circularly polarized source. The region constrained by W49B alone is markedly narrower in mass than the combined limit. Either omitting J2021+3651 from the pulsar ensemble or including only that source leads to more noticeable ``islands'' of enhanced detection probability surrounded by regions where a detectable signal is not expected. 

\bibliography{ALP-pol}

\end{document}